 \documentclass[final,1p]{elsarticle}

\usepackage{comment, color, graphicx,amsmath}

\usepackage{amssymb}

\journal{J. Comp. Phys.}

\newcommand{\onlinecite}{\cite}
\newcommand{\Deu}{\mathrm{D}}
\newcommand{\Mo}{\mathrm{Mo}}
\newcommand{\exponent}{k}

\newcommand{\Zeff}{Z_{\mathrm{eff}}}

\newcommand{\xmax}{x_{\mathrm{max}}}
\newcommand{\vv}{v}
\newcommand{\vth}{v_{\mathrm{th}}}
\newcommand{\erf}{\mathrm{erf}}

\newcommand{\ve}{\vv_{\mathrm{e}}}

\newcommand{\xe}{x_{\mathrm{e}}}

\newcommand{\gradv}{\nabla_{\vect{v}}}

\newcommand{\Leg}{\ell}
\newcommand{\xMax}{x_{\mathrm{Max}}}

\newcommand{\nue}{\nu_{\mathrm{e}}}

\newcommand{\nee}{n_{\mathrm{e}}}
\newcommand{\feo}{f_{\mathrm{e0}}}
\newcommand{\feone}{f_{\mathrm{e1}}}
\newcommand{\fione}{f_{\mathrm{i1}}}

\newcommand{\Te}{T_{\mathrm{e}}}
\newcommand{\Ti}{T_{\mathrm{i}}}

\newcommand{\pe}{p_{\mathrm{e}}}
\newcommand{\ppi}{p_{\mathrm{i}}}
\newcommand{\Lone}{\mathcal{L}_{31}}
\newcommand{\Ltwo}{\mathcal{L}_{32}}
\newcommand{\Lfour}{\mathcal{L}_{34}}

\newcommand{\Cee}{C_{\mathrm{ee}}}

\newcommand{\Cei}{C_{\mathrm{ei}}}

\renewcommand{\ni}{n_{\mathrm{i}}}

\newcommand{\me}{m_{\mathrm{e}}}
\newcommand{\mi}{m_{\mathrm{i}}}
\renewcommand{\ni}{n_{\mathrm{i}}}

\newcommand{\vpar}{\vv_{||}}

\newcommand{\vect}[1]{\mbox{\boldmath $#1$}}

\newcommand{\Lo}{\mathcal{L}}
\newcommand{\LoGyro}{\mathcal{L}_\varphi}

\newcommand{\be}{\begin{equation}}
\newcommand{\ee}{\end{equation}}
\newcommand{\PS}{Pfirsch-Schl\"{u}ter~}
\newcommand{\lla}{\left\langle}
\newcommand{\rra}{\right\rangle}

\newcommand{\vperp}{v_\bot}

\newcommand{\Cmod}{C_{\mathrm{mod}}}

\begin{document}

\begin{frontmatter}



\author{Matt Landreman\corref{Corresponding author}}
\ead{landrema@mit.edu}
\author{Darin R. Ernst}
\address{Massachusetts Institute of Technology, Plasma Science and Fusion Center}

\title{New velocity-space discretization for continuum kinetic calculations and Fokker-Planck collisions}


\author{}

\address{}

\begin{abstract}

Numerical techniques for discretization of velocity space
in continuum kinetic calculations are described.
An efficient spectral collocation method is developed for the speed coordinate -- the radius in velocity space --
employing a novel set of non-classical orthogonal polynomials.
For problems in which Fokker-Planck collisions are included,
a common situation in plasma physics,
a procedure is detailed to accurately and efficiently treat the field term
in the collision operator (in the absence of gyrokinetic corrections).
When species with disparate masses are included simultaneously,
a careful extrapolation of the Rosenbluth potentials
is performed.
The techniques are demonstrated in several applications,
including neoclassical calculations of the bootstrap current and plasma flows in a tokamak.

\end{abstract}


\end{frontmatter}


\section{Introduction}

A ubiquitous situation in numerical kinetic calculations is that
the distribution function must be discretized in a manner allowing
both accurate integration and accurate differentiation.
Integration is needed because typically moments of the distribution
function, such as density and velocity, are of interest.
Differentiation is needed both for the collisionless terms in the kinetic equation
and also for velocity-space diffusion in collisions.
Spherical or cylindrical coordinates are natural for velocity space,
meaning the normalized spherical or cylindrical radius (speed)
coordinate $x$ lies in the semi-infinite domain $[0,\infty)$.
The distribution function often has a Maxwellian envelope, meaning it behaves
as $\propto\exp(-x^2)$ as $x \to \infty$.

Many discretization schemes for the radial velocity coordinate are possible and many have
been explored in the literature\cite{Valentini, MichaelGrid, AstroGK, WongChan, Pataki, NEOFP, Lyons},
each with advantages and disadvantages
regarding the above requirements.
A uniform grid allows modest accuracy at both integration and differentiation
using finite difference/volume/element methods.
To use a uniform grid, $x$ may either be mapped to a finite interval using a coordinate transformation,
or else the fact that the distribution function is exponentially small for $x\gtrsim 6$ may
be used to truncate the $x$ grid above some $\xmax$.
Alternatively, Gaussian
abscissa permit extremely accurate integration but generally only low-order differentiation
if finite difference/volume/element methods are applied to the unequally spaced grid.
A Chebyshev grid permits both spectrally accurate differentiation and
integration\cite{ClenshawCurtis, Trefethen}. However, Chebyshev grids involve a high density
of nodes near the endpoints of a finite interval, so unless a transformation is
applied, Chebyshev grids are therefore poorly suited
for the semi-infinite domain of $x$ and $\propto \exp(-x^2)$ dependence typical of distribution functions.
Methods using Laguerre or associated Laguerre polynomials
have also been used, but as we will show, these methods
do not always perform as well as one might hope
due to a nonanalytic Jacobian in the coordinate transformation from energy to speed.

In this work, we present a new approach for discretizing velocity space.
The approach permits both spectrally accurate integration and differentiation
on the semi-infinite domain $[0, \infty)$ for functions
with Maxwellian-like $\propto \exp(-x^2)$ dependence
for large normalized speed $x$.
The method is collocation rather than modal in nature,
i.e. the function is known on a set of grid points (abscissae)
rather than being explicitly expanded in a set of modes.
As such, the method is well suited for nonlinear computations
in addition to linear ones.

Our method derives from a set of previously
unexplored orthogonal polynomials.
The semi-infinite integration domain in the orthogonality relation for these new polynomials
is identical to that of the Laguerre polynomials.
However, the use of a different weight $e^{-x^2}$ in place of $e^{-y}$ results
in polynomials with superior properties for the calculations
of interest.

In kinetic calculations for plasmas,
it is often important to accurately include collisions in the kinetic equation,
and the Fokker-Planck operator\cite{RMJ}
is the best available description of collisions.
This operator may be written in terms of ``Rosenbluth potentials",
which are defined in terms of the distribution function
through a pair of Poisson equations.
A complication of the Fokker-Planck operator is that the Rosenbluth
potentials vary as powers of $x$ rather than as $\exp(-x^2)$
for large $x$, which means
discretization schemes that work well for the distribution function
may not work well for the potentials.
Accurate numerical schemes for handling
the Fokker-Planck operator in plasma computations
have been a subject of great interest \cite{Pareschi, Xiong, Pataki, Spencer}.
Here we will develop an efficient approach to incorporating
the Rosenbluth potentials in the kinetic equation,
carefully accounting for their behavior at large $x$
to maintain high precision even for coarse grid resolution.

The new techniques we discuss are demonstrated in several
applications.  First, we compute the resistivity of a plasma.
Second, we compute the bootstrap current in a tokamak plasma.
Lastly, we calculate the flows of multiple ion species in a tokamak.
These computations require the solution of
drift-kinetic equations, equations in which both the collision operator and other kinetic
terms appear.  Using the new velocity discretization
described here,
we find that only 4-6 grid points in $x$ are required for the desired level of convergence.
For gyrokinetic simulations of plasma turbulence, which commonly use $\sim 16$
energy gridpoints, this new energy grid may reduce requirements of time, memory, or number of processors.

\section{Spectral collocation scheme for velocity space}
\label{sec:polynomials}

For a variety of problems in kinetic theory, either with or without collisions, it is useful
to represent the distribution function in either spherical or cylindrical coordinates
in velocity space.  The dimensional coordinates $v$ (the spherical radius in velocity space)
or $\vperp$ (the cylindrical radius in velocity space) then arise.
Either coordinate may be normalized for numerical work by the thermal speed $\vth = \sqrt{2T/m}$
where $T$ is a typical temperature and $m$ is the mass of the particle species.
We define $x = v/\vth$ or $x=\vperp/\vth$ as appropriate.

For large $x$, distribution functions decay exponentially as
$\propto \exp(-x^2)$.
It is therefore natural to represent the distribution function as a sum of polynomials $P_n^{\exponent}$
that are orthogonal according to the relevant weight and domain:
\begin{equation}
\int_0^\infty P_N^{\exponent} (x) P_n^{\exponent}(x) x^{\exponent} e^{-x^2}\, dx = M_n^{\exponent} \delta_{N,n}
\label{eq:orthogonality}
\end{equation}
where $\exponent$ is any number greater than $-1$, and $M_n^{\exponent}$ is some normalization.
Notice (\ref{eq:orthogonality}) differs from the defining orthogonality relations
for both the associated Laguerre and Hermite polynomials, which are, respectively,
\begin{equation}
\int_0^\infty L_N^m(y) L_n^m(y) y^m e^{-y}\, dy = \frac{\Gamma(n+m+1 )}{n!} \delta_{N,n}
\end{equation}
(i.e. polynomials in $x^2$ rather than $x$) and
\begin{equation}
\int_{-\infty}^\infty H_N(x) H_n(x) e^{-x^2}\, dx = 2^n n! \sqrt{\pi} \delta_{N,n}
\end{equation}
(different range of integration than (\ref{eq:orthogonality})).
Laguerre polynomials are the special case of associated Laguerre polynomials with $m=0$.
There are several reasons why it is preferable to use polynomials in speed $x$
rather than polynomials in energy $y=x^2$, i.e why the new polynomials
are preferable to Laguerre or associated Laguerre polynomials.
These reasons will be developed throughout the remainder of this section.
As initial motivation, consider that the new polynomials can represent both even and odd powers of $v$ or $\vperp$,
whereas the associated Laguerre polynomials can represent only
even powers.

In our experience, the choice $\exponent=0$ tends to yield the fastest convergence
for the problems we consider in the following sections,
so for the rest of this paper we consider the polynomials $P_n = P_n^0$.
It is straightforward to generalize all results and algorithms presented below
to the case of different $\exponent$.

The first few polynomials may be computed iteratively using
the following Gram-Schmidt procedure (though
this method turns out to be poorly conditioned when $n$ is large).
The polynomial $P_n(x)$ has $n+1$ coefficients, which may be determined by imposing orthogonality with respect
to $P_0$ through $P_{n-1}$ and enforcing the normalization, for a total of $n+1$ constraints.
The first few polynomials $P_n$, normalized so the leading coefficient is 1, are thus
\begin{eqnarray}
P_0(x) &=& 1 \label{eq:monicPolynomials}\\
P_1(x) &=& x-\frac{1}{\sqrt{\pi}} \nonumber \\
P_2(x) &=& x^2 -\frac{\sqrt{\pi}}{\pi-2}x +\frac{4-\pi}{2(\pi-2)} \nonumber \\
P_3(x) &=& x^3
- \frac{3\pi-8}{2\sqrt{\pi}(\pi-3)} x^2
+ \frac{10-3\pi}{2(\pi-3)}x
- \frac{16-5\pi}{4\sqrt{\pi}(\pi-3)}. \nonumber
\end{eqnarray}
The same polynomials normalized so $M_n^0=1$ are
\begin{eqnarray}
P_0(x) &=& \frac{\sqrt{2}}{\pi^{1/4}} = 1.0623 \label{eq:normalizedPolynomials}\\
P_1(x) &=& 2.4921x-1.4060 \nonumber \\
P_2(x) &=& 4.2656 x^2 -6.6228 x + 1.6037 \nonumber \\
P_3(x) &=& 6.0027 x^3 - 17.0392 x^2 + 12.1931 x - 1.7463. \nonumber
\end{eqnarray}
The appearances of $\sqrt{\pi}$ and $\pi$ in (\ref{eq:monicPolynomials})
highlight that
the $P_n$ polynomials do not seem to be simply related to any of the classical orthogonal polynomials.

\begin{figure}
\includegraphics[width=\textwidth]{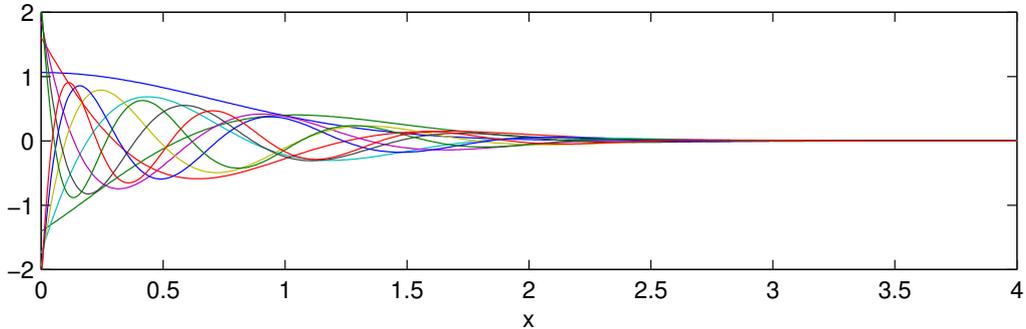}
\caption{(Color online)
The first 10 new polynomial modes $P_n(x) \exp(-x^2)$, normalized so $M_n^0=1$.
\label{fig:newPolynomials}}
\end{figure}

\begin{figure}
\includegraphics[width=\textwidth]{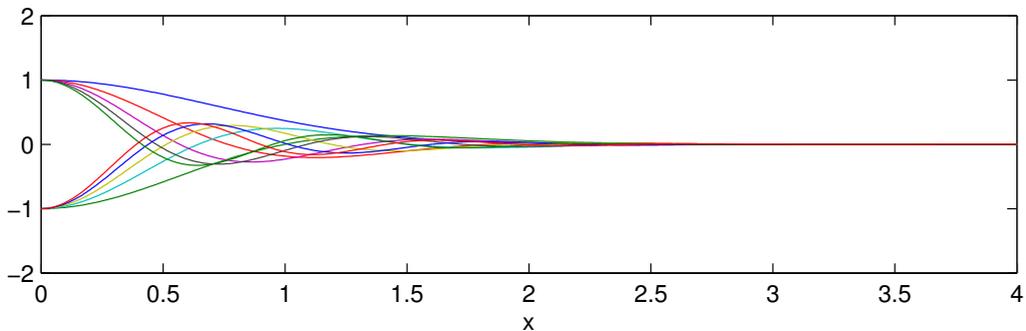}
\caption{(Color online)
The first 10 Laguerre modes $L_n(x^2) \exp(-x^2)$ (normalized).
\label{fig:Laguerres}}
\end{figure}

\begin{figure}
\includegraphics[width=\textwidth]{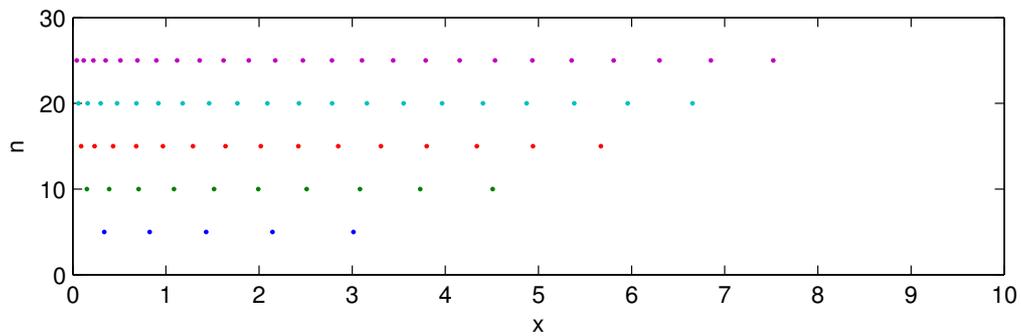}
\caption{(Color online)
Zeros of the new polynomials $P_n(x)$, i.e. the speed grid for Gaussian integration
and for the new spectral collocation method.
\label{fig:newPolynomialZeros}}
\end{figure}

\begin{figure}
\includegraphics[width=\textwidth]{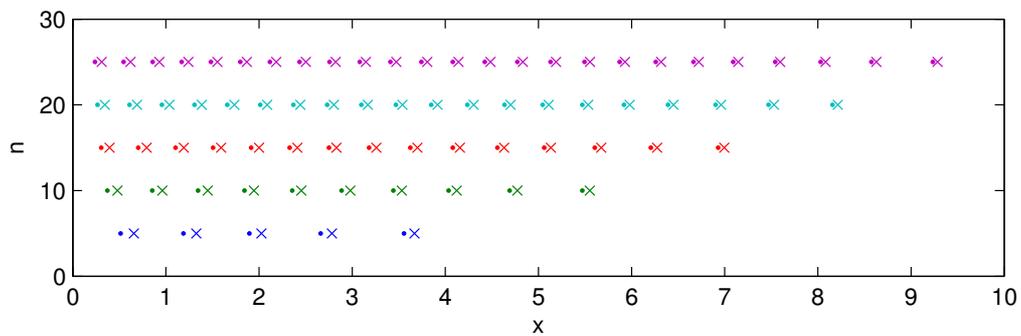}
\caption{(Color online)
Values of $x=\sqrt{y}$ for zeros of the Laguerre polynomials $L_n(y)$ (dots) and associated Laguerre polynomials $L_n^{(1/2)}(y)$ (crosses),
i.e. the speed grids for Gaussian integration and for spectral collocation methods.
\label{fig:LaguerreZeros}}
\end{figure}

Figure \ref{fig:newPolynomials} shows
the first few normalized polynomials, multiplied by $e^{-x^2}$.
Figure \ref{fig:Laguerres} shows the
corresponding plot for the Laguerre polynomials.
(Associated Laguerre polynomials look very similar.)
These figures illustrate the first reason the new polynomials
are preferable:
they have structure localized more in the region
where the particle density is high ($x \lesssim 2$) compared to the $L^m_n(x^2) \exp(-x^2)$ modes.
For example, the new polynomial modes
have significant differences from each other in the range $x<1$
-- so they can accurately resolve structure in this region
where many particles reside --
whereas the Laguerre modes
are very similar to each other in the same range.
Even with 10 Laguerre modes, no structure
can be resolved within $x<0.5$ since there are no nodes there,
whereas the new modes permit ample resolution of this region.
The same issue can be seen in figures \ref{fig:newPolynomialZeros}
and \ref{fig:LaguerreZeros},
which show the grids for the corresponding spectral collocation methods.
In other words, the zeros of the $P_n(x)$ are plotted in Figure \ref{fig:newPolynomialZeros},
and the square roots of the zeros of the $L_n^{(m)}(y)$
are plotted in Figure \ref{fig:LaguerreZeros} to
show the grid points in
speed rather than energy.
The Laguerre and associated Laguerre bases tends to ``waste"
more nodes in the region $x > 2$ where the distribution function is nearly zero.

It should be mentioned that the abscissae can be scaled by a factor
slightly less than 1 without destroying the spectral convergence
properties while achieving better resolution at small values of $x$.

The clustering of nodes for the new
polynomials near $x=0$ turns out to be a major advantage when multiple particle species are considered simultaneously,
discussed in section \ref{sec:impurities}, since high resolution is needed in this region when the species masses are disparate.

As with any set of orthogonal polynomials, the set $P_n$ is associated with a Gaussian
integration scheme, with the abscissae corresponding to the zeros of the polynomials.
These abscissae and Gaussian integration weights may be computed
for any orthogonality relation
using the ``Stieltjes procedure",
an alternative method to construct the polynomials which is better conditioned
numerically than the direct orthogonalization approach described above.
The theory underlying the Stieltjes procedure is detailed in Refs. \onlinecite{Gautschi1, Gautschi2, Gautschi3},
with Fortran 77 subroutines implementing the algorithms in Ref. \onlinecite{ORTHPOL}.
Another practical implementation of the algorithm may be found in
Ref. \onlinecite{nrWebnote},
with documentation given in section 4.6
of Ref. \onlinecite{numericalRecipes}.
In brief, the algorithm involves first iteratively computing the coefficients $a_j$ and $b_j$
of the recurrence relation for the polynomials,
\begin{equation}
P_{j+1}(x) = (x-a_j)P_j(x) - b_j P_{j-1}(x),
\label{eq:recurrence}
\end{equation}
which may be done by evaluating
\begin{eqnarray}
a_j &=& \frac{\int_0^{\infty} P_j^2(x) x\, e^{-x^2}dx}{\int_0^{\infty} P_j^2(x)  e^{-x^2}dx } \label{eq:aj}\\
b_j &=& \frac{\int_0^{\infty} P_j^2(x) e^{-x^2}dx}{\int_0^{\infty} P_{j-1}^2(x)  e^{-x^2}dx } \label{eq:bj}.
\end{eqnarray}
The integrals in (\ref{eq:aj})-(\ref{eq:bj}) are performed using adaptive quadrature, evaluating
$P_j(x)$ using the recurrence relation.
The resulting recurrence coefficients are then used to form the tridiagonal
Jacobi matrix:
\begin{equation}
\vect{J} = \left(
\begin{array}{ccccc}
a_0 & \sqrt{b_1} &&& \\
\sqrt{b_1} & a_1 & \sqrt{b_2} && \\
& \ddots & \ddots & \ddots & \\
 && \sqrt{b_{n-2}} & a_{n-2} & \sqrt{b_{n-1}} \\
 &&& \sqrt{b_{n-1}} & a_{n-1} \\
\end{array}
\right).
\end{equation}
As shown in the aforementioned references, the eigenvalues of this matrix
give the abscissae.  Furthermore, the first elements of the normalized eigenvectors may
be squared and multiplied by $\int _0^{\infty} e^{-x^2}dx = \sqrt{\pi}/2$ to obtain the integration weights.
This algorithm is numerically well conditioned, and
produces a grid with extremely accurate integration properties, as
we will demonstrate shortly.

The above procedure produces a grid with no point at $x=0$.
If a point is needed there to impose a boundary condition,
i.e. if a Gauss-Radau grid is desired,
the bottom-right element of the Jacobi
matrix may be replaced by $-b_{n-1} P_{n-2}(0)/P_{n-1}(0)$,
as discussed in Section 6.2 of Ref. \onlinecite{Gautschi3}.
The abscissae and weights are then computed from the eigenvalues and eigenvectors as before.
The resulting grid will still be extremely accurate for integration,
but slightly less so
than the grid with no point at $x=0$.

As mentioned previously, for kinetic calculations we need not only an accurate
integration scheme for the grid, but also an accurate differentiation matrix.
The spectral differentiation matrix for any set of
nodes and any weighting function may be computed using the \emph{poldif}
algorithm described in Refs. \onlinecite{DMSuite, DMSuite2, DMSuite3}.
The algorithm amounts to representing a function
on the grid as a weighted sum of the polynomials,
differentiating the polynomials,
and evaluating the result on the grid.
We apply the algorithm using the weight $e^{-x^2}$,
and using the abscissae calculated from Stieljes procedure above.

\begin{figure}
\centering
\includegraphics[height=7in]{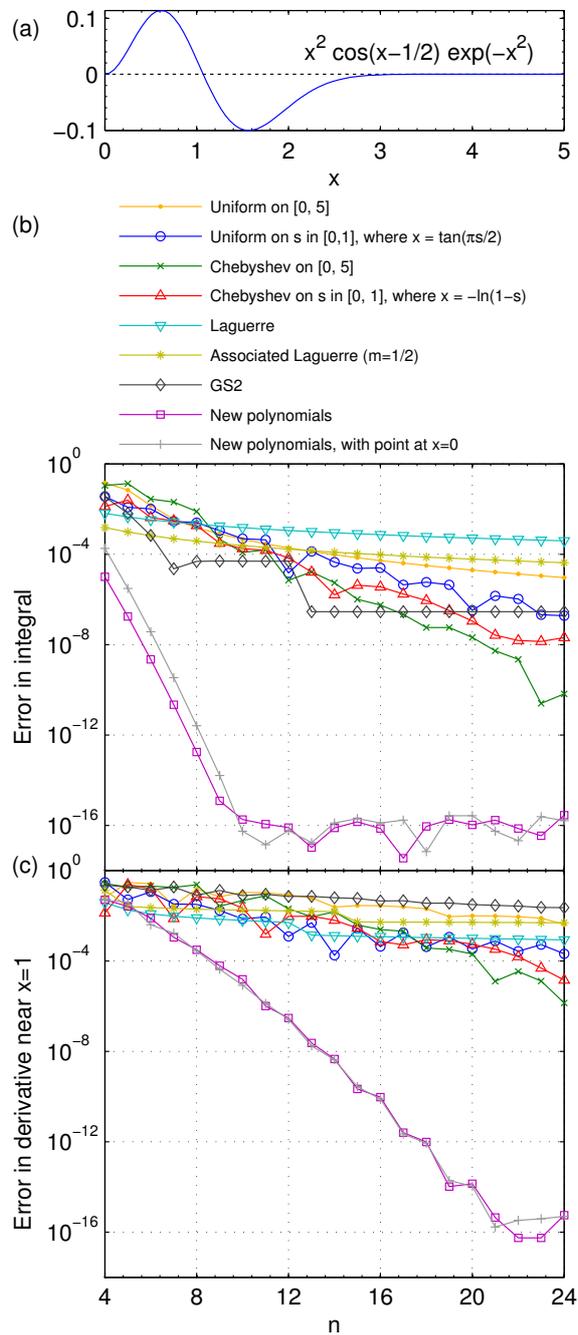}
\caption{(Color online)
A central result of this paper:
for the function in (a),
typical of the velocity dependence of a distribution function,
the new polynomial spectral collocation method
described here far outperforms other schemes
for both (b) integration and (c) differentiation.
In (b)-(c), $n$ denotes the number of grid points.
\label{fig:grids1}}
\end{figure}

\begin{figure}
\centering
\includegraphics[height=7in]{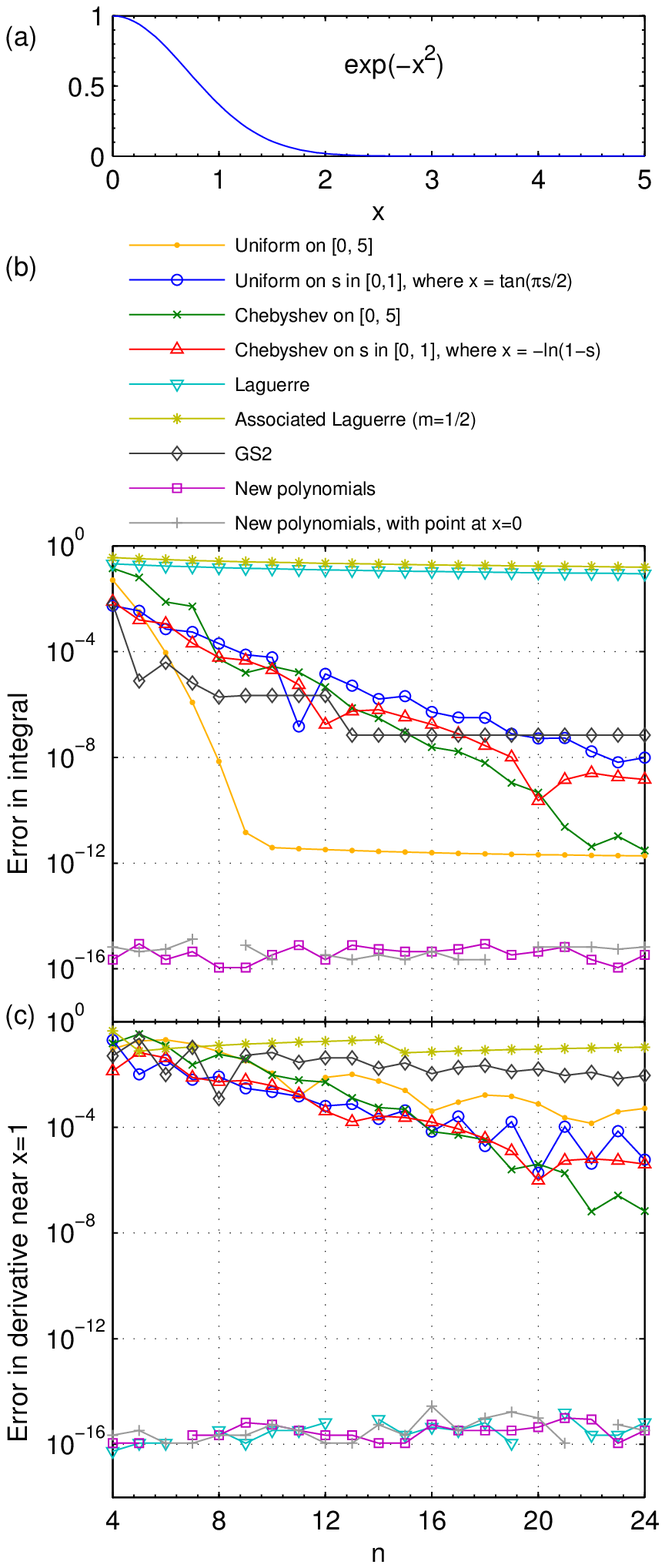}
\caption{(Color online)
Same as Figure \ref{fig:grids1},
but for the function in (a).
Missing points indicate zero error to machine precision.
\label{fig:grids3}}
\end{figure}

\begin{figure}
\centering
\includegraphics[height=7in]{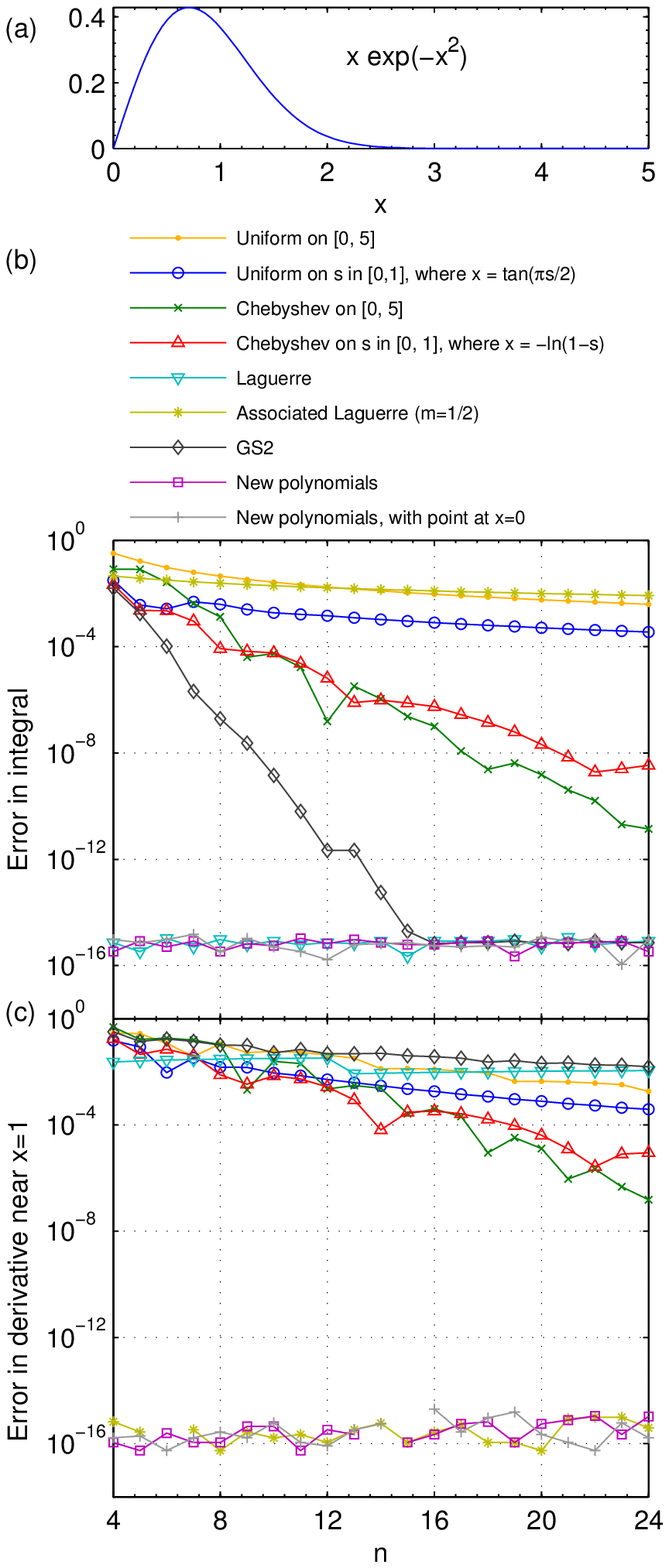}
\caption{(Color online)
Same as Figure \ref{fig:grids1},
but for the function in (a).
Missing points indicate zero error to machine precision.
\label{fig:grids4}}
\end{figure}

\begin{figure}
\centering
\includegraphics[height=7in]{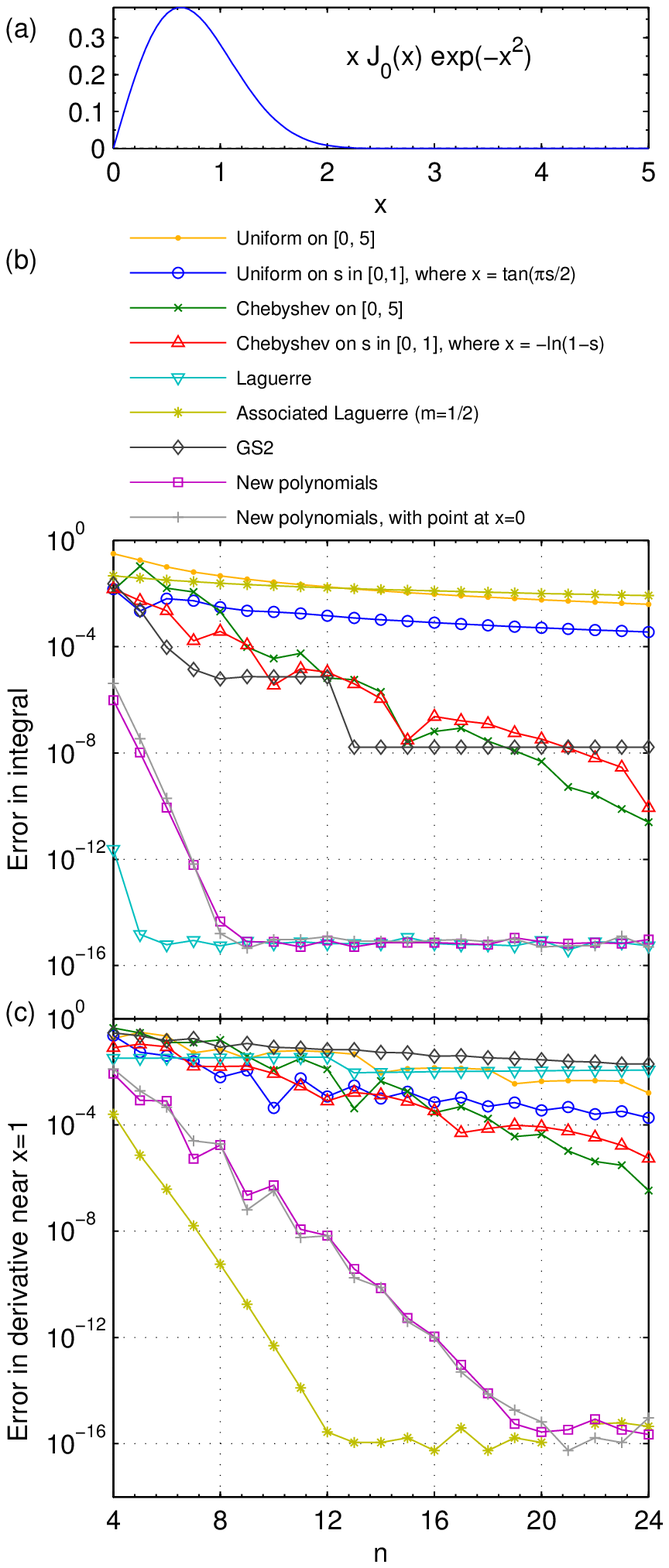}
\caption{(Color online)
Same as Figure \ref{fig:grids1},
but for the function in (a).
\label{fig:grids45}}
\end{figure}

\begin{figure}
\centering
\includegraphics[height=7in]{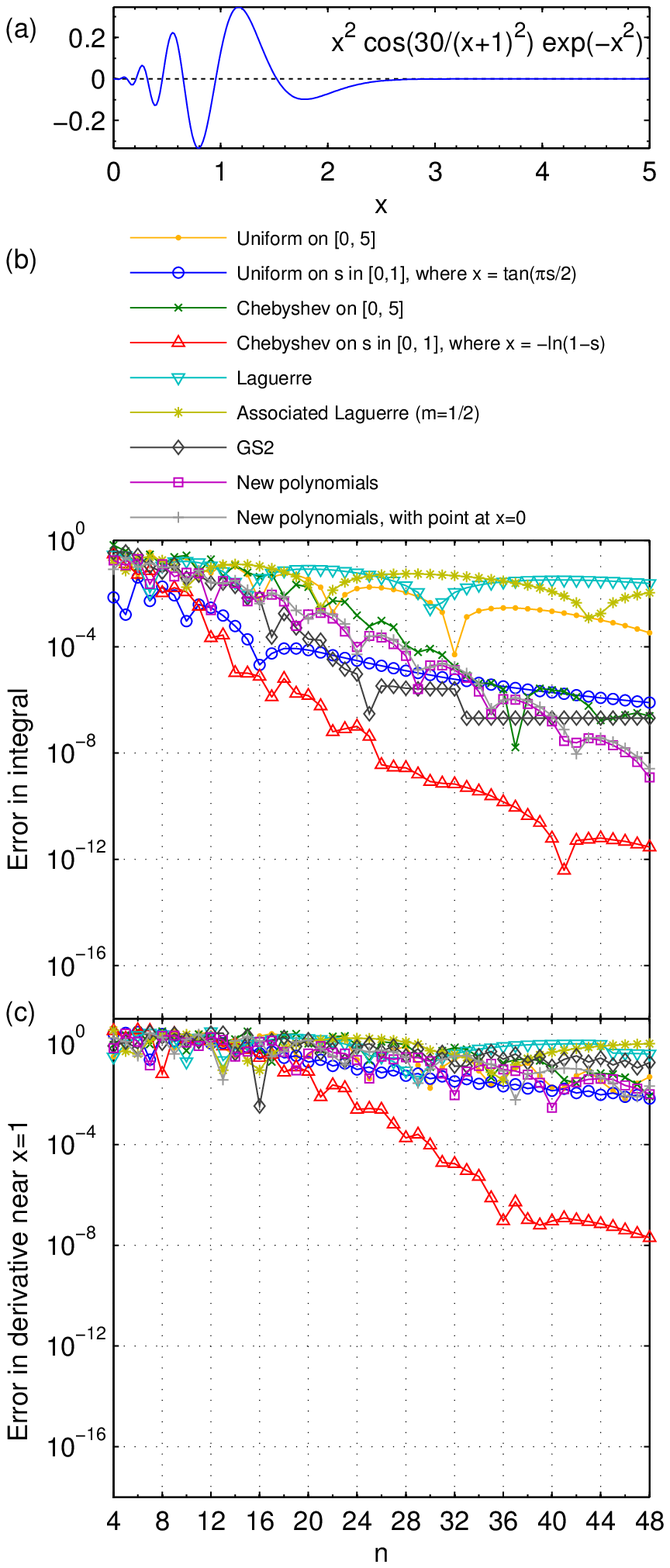}
\caption{(Color online)
Same as Figure \ref{fig:grids1},
but for the function in (a).
\label{fig:grids5}}
\end{figure}

To demonstrate the efficacy of the new collocation method,
figures \ref{fig:grids1}-\ref{fig:grids5} compare
various discretization methods
for both integration and differentiation of several functions.
We first choose the function $x^2 \cos(x-1/2)\exp(-x^2)$
because 1) for large $x$
it has a Maxwellian-like envelope, 2) for small $x$ it has the $x^2$ behavior
typical of physical velocity moment integrands near $x=0$, and 3)
due to the $\cos(x-1/2)$ factor it cannot be
exactly represented by any of the polynomial approximation methods, and its Taylor series contains both even and odd powers of $x$.
We measure the accuracy of differentiation by
comparing the numerical derivative to the analytical derivative at whichever grid
point is closest to 1.
For the new method, results are shown for both the standard Gauss abscissae (no point at $x=0$)
and for Gauss-Radau points (including $x=0$).
It can be seen that the new schemes perform substantially better
in both integration and differentiation than other methods.
The same results are reported in tabular form in Tables \ref{tab:integral}-\ref{tab:derivative}.
We have performed a similar analysis to many functions
besides that in figure \ref{fig:grids1}.a,
and the behavior seen in \ref{fig:grids1}.b-c is typical.

The relatively poor performance of both the Laguerre and associated Laguerre methods
in this comparison
may be understood as follows\cite{MichaelGrid,AstroGK}.  These grids can
allow very accurate integration and differentiation,
but only for functions that are analytic in $y=x^2$,
the argument of the polynomials in the orthogonality relation that defines them.
Since $x=\sqrt{y}$ is nonanalytic in $y$ at $y=0$,
the accuracy of Laguerre and associated Laguerre methods
fails for general functions of $x$.
This issue has a practical implication for kinetic computations.
In kinetic codes,
it is common to want both the density moment $(\int d^3v\, f \propto \int_0^{\infty} dx\, x^2 f = \frac{1}{2}\int_0^{\infty} dy\sqrt{y} f)$
and velocity moment $(\int d^3v\, \vect{v}f \propto \int_0^{\infty} dx\, x^3 f = \frac{1}{2}\int_0^{\infty} dy\, y f)$
of the distribution function $f$.  Even if $f$ is analytic in $y$,
the spectral accuracy of Gaussian integration is lost when computing
the density moment using a Laguerre grid due to the $\sqrt{y}$ factor, as noted in Refs. \onlinecite{MichaelGrid,AstroGK}.
This problem could be avoided using an associated Laguerre grid with $m=1/2$,
so the nonanalytic factor $\sqrt{y}$ is incorporated into the integration weight.
However, then accuracy would be lost for the velocity integrand, since it includes an additional factor of $\sqrt{y}$.
By using polynomials in $x$ rather than polynomials in $y$, good accuracy can be retained for
both integrals simultaneously.

The plots also include results for a uniform grid and Chebyshev grid covering the interval $[0,5]$.
We choose $x=5$ as a reasonable cutoff
since the model function is exponentially small ($<3\times 10^{-10}$) for $x>5$.
For the uniform grid, integration is performed using
the trapezoid rule, and differentiation
is performed using 5-point finite differencing.

Also shown are results for several schemes in which
$x\in [0,\infty)$ is mapped to a new variable $s$ in the finite interval $[0,1]$.
We considered the three maps $s = -\ln(1-x)$, $x=\tan(\pi s/2)$,
and $x=s/(1-s)$, applying either a uniform or Chebyshev grid to $s$.
Results are shown for the map which produced the most accurate results.
The direct Chebyshev grids in $x$ and $s$ perform roughly
equally well to each other, while the uniform grid in $s$ tends to outperform
the uniform grid in $x$.

Results are also plotted for the collocation scheme
used in the gyrokinetic code GS2, detailed in Ref. \onlinecite{MichaelGrid}.
The same grid is also used in the astrophysics code AstroGK\cite{AstroGK}.
In this scheme, the grid consists of $n-m$ Gauss-Legendre points on the interval
$[0, 2.5]$, together with $m$ Gauss-Laguerre points on the interval $[2.5, \infty)$,
where $m=1$ for $n\le12$ and $m=2$ otherwise.
The philosophy behind this approach is to only use the Gauss-Laguerre points
away from the singularity at $x=0$.
Indeed, the GS2 approach shows substantially better convergence of the integral than the Laguerre-only
or associated Laguerre methods.
The actual GS2 fortran subroutines are used to generate both the abscissae and weights
for the figures here.
Convergence of the integral stagnates around $10^{-7}-10^{-8}$
since no more than 2 points are ever used in the interval $[2.5,\infty)$; if
$m$ is increased beyond 2, the GS2 method can reach machine precision.
The
derivative is computed using a 3-point finite-difference stencil,
just as is used for the second derivative in GS2.
This low-order scheme was used in GS2 because it allowed exact numerical
conservation of particles in velocity space, which higher-order methods
generally do not allow.

\begin{table}
\centering
\begin{tabular}{l c c c}
Discretization scheme & $n=4$ & $n=8$ & $n=16$ \\
\hline
Uniform on $[0,5]$ & $1\times 10^{-1}$ & $\;\;1\times 10^{-3}\;\;$ & $5\times 10^{-5}$ \\
Uniform in $s$, $x=\tan(\pi s/2)$ & $3\times 10^{-2}$ & $2\times 10^{-3}$ & $2\times 10^{-5}$ \\
Chebyshev on $[0,5]$ & $1\times 10^{-1}$ & $8\times 10^{-3}$ & $6\times 10^{-7}$ \\
Chebyshev in $s$, $x=-\ln(1-s)$ & $1\times 10^{-2}$ & $2\times 10^{-3}$ & $4\times 10^{-6}$ \\
Laguerre & $6\times 10^{-3}$ & $2\times 10^{-3}$ & $7\times 10^{-4}$ \\
Associated Laguerre ($m=1/2$) & $2\times 10^{-3}$ & $4\times 10^{-4}$ & $9\times 10^{-5}$ \\
GS2 & $3\times 10^{-2}$ & $5\times 10^{-5}$ & $3\times 10^{-7}$ \\
New polynomials & $1\times 10^{-5}$ & $2\times 10^{-13}$ & $7\times 10^{-17}$ \\
New polynomials with $x=0$ & $2\times 10^{-4}$ & $3\times 10^{-12}$ & $1\times 10^{-16}$ \\
\end{tabular}
\caption{Error in the integral of $x^2 \cos(x-1/2) e^{-x^2}$
for various numbers of grid points $n$.
\label{tab:integral}}
\end{table}

\begin{table}
\centering
\begin{tabular}{l c c c}
Discretization scheme & $n=4$ & $n=8$ & $n=16$ \\
\hline
Uniform on $[0,5]$ & $6\times 10^{-2}$ & $\;\;2\times 10^{-1}\;\;$ & $3\times 10^{-2}$ \\
Uniform in $s$, $x=\tan(\pi s/2)$ & $3\times 10^{-1}$ & $3\times 10^{-2}$ & $4\times 10^{-4}$ \\
Chebyshev on $[0,5]$ & $3\times 10^{-1}$ & $2\times 10^{-1}$ & $2\times 10^{-3}$ \\
Chebyshev in $s$, $x=-\ln(1-s)$ & $1\times 10^{-2}$ & $7\times 10^{-2}$ & $7\times 10^{-4}$ \\
Laguerre & $4\times 10^{-2}$ & $8\times 10^{-3}$ & $1\times 10^{-3}$ \\
Associated Laguerre ($m=1/2$) & $1\times 10^{-1}$ & $2\times 10^{-2}$ & $5\times 10^{-3}$ \\
GS2 & $2\times 10^{-1}$ & $8\times 10^{-2}$ & $5\times 10^{-2}$ \\
New polynomials & $5\times 10^{-2}$ & $3\times 10^{-4}$ & $9\times 10^{-11}$ \\
New polynomials with $x=0$& $5\times 10^{-2}$ & $3\times 10^{-4}$ & $8\times 10^{-11}$ \\
\end{tabular}
\caption{Error in the derivative of $x^2 \cos(x-1/2) e^{-x^2}$
for various numbers of grid points $n$,
reported at the grid point closest to $x=1$.
\label{tab:derivative}}
\end{table}

Figures \ref{fig:grids3}-\ref{fig:grids4}
show interesting effects arising when the various integration
and differentiation schemes are applied to $x^a e^{-x^2}$
for various integers $a$.  The new polynomial scheme gives
answers correct to machine precision both for integration (when $a<2n$)
and for differentiation (when $a<n$).
The Laguerre or associated Laguerre grids with integer $m$
can get the derivative correct to machine precision when $a$ is even,
or they can compute the integral correct to machine precision when $a$ is odd,
but they perform poorly in the opposite cases.
This behavior follows because the Laguerre method is designed to
exactly evaluate $\int_0^\infty dy\, y^j e^{-y} = 2\int_0^\infty dx\, x^{2j+1}e^{-x^2}$
and to exactly differentiate $y^j e^{-y}$ for integers $j$,
but they will perform poorly when the nonanalytic factor $\sqrt{y}$
is included in the integral or derivative.
Conversely, associated Laguerre grids with half-integer $m$
can compute the integral correct to machine precision when $a$ is even,
or they can compute the derivative correct to machine precision when $a$ is odd,
but they perform rather poorly in the opposite cases.
The new polynomials get all cases correct to machine precision for
both even and odd $a$.
In figure (\ref{fig:grids4}).b,
the GS2 grid can reach machine precision for integration
of this special integrand because on the interval $[2.5,\infty)$, the function can be exactly integrated
by the Laguerre grid, even with just 2 points.  Convergence
is therefore limited only by convergence of the Legendre grid on the interval $[0,2.5]$.

The integration and differentiation properties of the Laguerre and associated Laguerre grids
on $x^a e^{-x^2}$ for even and odd $a$ extend in part to all even and odd functions of $x$.
The latter is illustrated in figure \ref{fig:grids45}, which shows the performance of the various grids
for the odd function $j(x) = x J_0(x)\exp(-x^2)$, where $J_0$ is a Bessel function.
(We choose $j(x)$ since functions like it arise
in gyrokinetics.)  As $x J_0(x)$ may be expanded in a Taylor series about $x=0$
with only odd powers of $x$, $j(x)$ is integrated accurately on a Laguerre grid,
but not differentiated accurately on this grid.
Similarly, $j(x)$ is differentiated accurately on an associated Laguerre grid with half-integer $m$,
but it is not integrated accurately on this grid.
If the $x$ preceding $J_0$ is replaced by an even power of $x$,
the behaviors of integer and half-integer associated Laguerre grids are reversed.
However, as shown in figure \ref{fig:grids45}, the new polynomial grids allow very accurate integration and differentiation simultaneously.

We do find that for some functions with small-scale oscillatory structure,
the new polynomials may not be the most accurate discretization option.  Such a function is illustrated
in figure \ref{fig:grids5}.  For this function,
all grid types show slower
convergence of both the integral and derivative
compared to previous functions. (Note the expanded $n$ axis
in figure (\ref{fig:grids5}).)
The Chebyshev grid in the remapped variable performs best,
and the uniform remapped grid also performs well.
The superior performance of these grids can be understood
in part from their very high density of points near $x=0$,
higher even than the new polynomials,
helping to resolve the fine structure in this region.
For problems in which fine structure is expected in the $x$ coordinate,
such as gyrokinetic turbulence,
it may be useful to build codes that can easily switch between several types of grids,
since it may not be clear \emph{a priori} which will actually yield faster convergence in practice for the specific problem.

Let us now make a few remarks concerning conservation.
Exact conservation properties in spectral velocity discretization
schemes have been examined previously in Ref. \cite{Halloway}
for the one-dimensional collisionless Vlasov equation,
containing advective terms with no diffusion.
In that work, the velocity coordinate has domain $(-\infty,\infty)$, instead of $[0,\infty)$ as in the present work.

Consider a kinetic equation
which conserves mass
and which contains a diffusive term $v^{-2} \partial \Gamma/\partial v$
for some $\Gamma$.
When the $x=v/\vth$ coordinate in this kinetic equation
is discretized,
mass conservation is preserved if
the relation
\begin{equation}
\int _0^\infty dx (dA/dx)=-A(0)
\label{eq:ibp}
\end{equation}
holds exactly for any
suitable function $A(x)$ using the discrete integration and differentiation operators.
The version of the new polynomial discretization scheme with a point at $x=0$ (Gauss-Radau)
exactly satisfies this relation, whereas the new scheme without a point at $x=0$
does not satisfy (\ref{eq:ibp}) exactly. To see why, consider that applying the discretized $d/dx$
operator returns the derivative of a polynomial approximation of $A(x)$,
and the discretized $\int_0^\infty$ operator then returns the exact integral of the resulting polynomial.
Thus, the discrete  $\int _0^\infty (d/dx)$ operation returns the value of the polynomial approximation of $A$  at $x=0$.
When there is a grid point located at $x=0$, the polynomial approximation to $A$
is exact there, whereas when there is no point at $x=0$, there will generally be some discrepancy.

Although it is therefore possible to exactly conserve mass in the Gauss-Radau version of our
discretization,
momentum and energy may or may not be conserved depending on other terms in the kinetic equation.
Conservation of momentum and energy can hold in certain kinetic problems (prior to discretization) due to
a balance between diffusive terms and other ``field particle'' collision terms,
which we will write explicitly in section \ref{sec:potentials}.
Upon discretization, this balance will generally be slightly altered so that momentum
and energy conservation is no longer exact. However,
Ref. \onlinecite{MichaelCollisionOp} has shown how momentum and energy conservation
can be maintained for certain
model collision operators
as long as the discrete integration and differentiation schemes satisfy (\ref{eq:ibp}).
Thus, replacing the integration and differentiation rules in Ref. \onlinecite{MichaelCollisionOp}
with our Gauss-Radau scheme would give exact momentum and energy conservation.

Perfect conservation may be required in calculations that advance time, since non-conservation errors can accumulate over time to cause larger qualitative changes. However, perfect conservation is not needed in time-independent equations or direct solution of eigenproblems, since non-conservation errors are then no worse than other discretization errors. The applications we consider throughout this paper are time-independent problems in which perfect conservation is unnecessary. However, conservation will likely be important in any future attempt to apply our
discretization methods to nonlinear time-dependent turbulence simulations.
In general, if any quantity is conserved in a kinetic equation prior to discretization,
any non-conservation errors become exponentially small using our methods
as the number of grid points in $x$ is increased.

Finally, we mention several possible variants of our discretization method.
First, the method could be recast as a modal expansion in the new polynomials,
with no grid.
Second, if a sparser differentiation matrix is necessary, a hybrid approach
could be used: the grid points and integration weights could be generated
as described following (\ref{eq:recurrence}), but differentiation could be performed using
a finite difference, finite volume, or finite element scheme on the irregular grid.
In finite difference/volume/element schemes, derivatives are computed using only information
from nearby grid points, in contrast to spectral collocation methods like ours,
in which derivatives are computed using information from every grid point.
Thus, the differentiation matrix in this hybrid scheme would be sparser
but less accurate; integration would retain high accuracy.

\section{Application 1: Resistivity}
\label{sec:resistivity}

As a first example of a physics problem in which the new grid is effective,
we now compute the resistivity of a plasma.
This problem requires both numerical integration and numerical differentiation of a Maxwellian-like function.
To avoid the complexity
of the Fokker-Planck collision operator,
which we will address in the next section, here we use a model for velocity-space
diffusion.
We consider the plasma to be unmagnetized and to consist of fully ionized hydrogen.
The resistivity is computed by solving the following electron kinetic equation\cite{HintonHazeltine, PerBook}
for the perturbed electron distribution function $\feone$:
\begin{equation}
\frac{eE}{\Te}\vpar \feo = \Cmod \{\feone\}
\label{eq:resistivityKE}
\end{equation}
where $e$ is the proton charge, $E$ is the electric field,
$\me$ is the electron mass, $\feo=\nee [\pi^{3/2} \ve^3]^{-1} \exp(-\xe^2)$
is a Maxwellian distribution function, $\nee$ is the electron density,
$\ve=\sqrt{2\Te/\me}$ is the electron thermal speed,
$\Te$ is the electron temperature, $\xe = v/\ve$,
and $\vpar = \vect{v}\cdot\vect{E}/|\vect{E}|$ is
the component of the velocity along the electric field.
The model collision operator $\Cmod$ we will use is
\begin{equation}
\Cmod\{\feone\} = \nue \frac{3\sqrt{\pi}}{4} \left[ \frac{Z + \erf(\xe)-\Psi}{v^3} \Lo\{\feone\}
+ \frac{1}{v^2}\frac{\partial}{\partial v}  \left(
2 \Psi \xe^2 \feone
+ \Psi v \frac{\partial \feone}{\partial v} \right)
\label{eq:Cmod}
\right]
\end{equation}
where $\nue = 4\sqrt{2\pi}\nee e^4\ln\Lambda/(3\sqrt{\me} \Te^{3/2})$
is the electron collision frequency,
$\ln\Lambda$ is the Coulomb logarithm, $Z=1$ is the ion charge,
$\Psi(\xe) = \left[ \erf(\xe) - 2 \pi^{-1/2} \xe e^{-\xe^2}\right]/(2\xe^2)$ is the Chandrasekhar function,
$\erf(\xe) = 2 \pi^{-1/2} \int_0^{\xe} e^{-w^2}dw$ is the error function,
\begin{equation}
\Lo = \frac{1}{2}\frac{\partial}{\partial\xi}(1-\xi^2)\frac{\partial}{\partial\xi},
\end{equation}
and $\xi=\vpar/\vv$.
The operator (\ref{eq:Cmod}), which includes diffusion in both pitch angle and energy,
is identical to the test particle part of the
electron-electron plus electron-ion Fokker-Planck operators
that will be discussed in the next
section and following (\ref{eq:fsa2}), and which can be derived from first principles.
Models such as (\ref{eq:Cmod}) are routinely used
in both analytic and numerical calculations\cite{HirshmanSigmar, MichaelCollisionOp}
since they are simpler than the full Fokker-Planck operator.

Upon solution of (\ref{eq:resistivityKE}) for $\feone$,
the resistivity $\eta$ is computed
from $\eta = E/j$ where $j=-e\int d^3v \; \vpar \feone$
is the current. Applying the ansatz $\feone(x,\xi) = \xi F(x)$ for some unknown $F(x)$ in (\ref{eq:resistivityKE})
and noting $\Lo \xi=-\xi$, $\xi$ may be divided out of the equation, leaving
only a one-dimensional problem in $x$.
To facilitate numerical computation, we define a dimensionless resistivity $\hat\eta = \eta \nee e^2/(\nue \me)$,
which can be calculated without specifying any of the physical constants.

\begin{figure}
\centering
\includegraphics{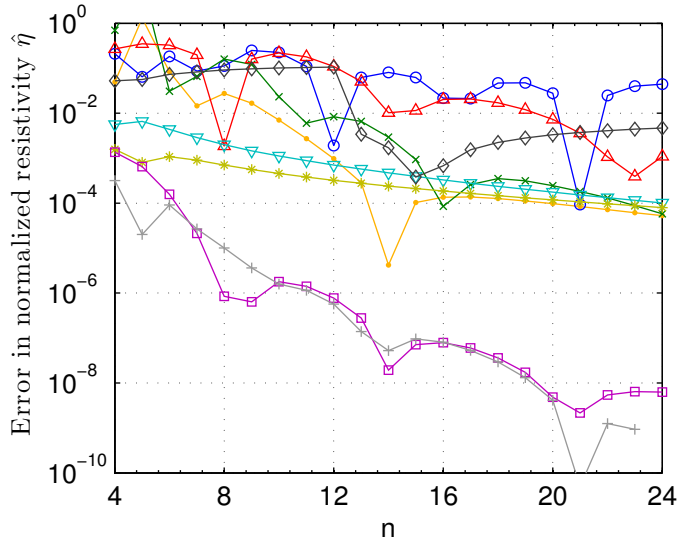}
\caption{(Color online)
Convergence in the resistivity of a hydrogen plasma
computed using various discretization methods,
and retaining only the test-particle part of the collision operator
for simplicity.
The curve for each discretization scheme is
labeled by the same
color and marker as in figures (\ref{fig:grids1})-(\ref{fig:grids5}).
\label{fig:resistivity}}
\end{figure}

Figure (\ref{fig:resistivity})
shows the departure of $\hat\eta$ from its converged value
as a function of grid size for the various grids considered earlier.
(The converged value, $\hat\eta\approx 0.99651105$, is roughly twice the classic Spitzer-H\"arm value\cite{Spitzer} $\hat\eta\approx 0.50611832$
which can obtained using
the improved collision operator we discuss in the next section.)
For these calculations, in schemes with a grid point at $\xe=0$, we impose $F=0$ there;
otherwise we do not impose a boundary condition at $\xe=0$.
For the uniform and Chebyshev grids on $[0,5]$,
we impose $F=0$ at $\xe=5$.
For the remapped uniform and Chebyshev grids, we impose $F=0$
at $\xe=\infty$.
For the Laguerre, associated Laguerre, and new polynomial
methods, no boundary condition is explicitly imposed at large $\xe$,
since the $\exp(-\xe^2)$ behavior is implicitly built into the basis functions for these methods.
As can be seen in figure (\ref{fig:resistivity}),
the new discretization scheme leads to substantially faster convergence than other discretization schemes.

\begin{figure}
\centering
\includegraphics{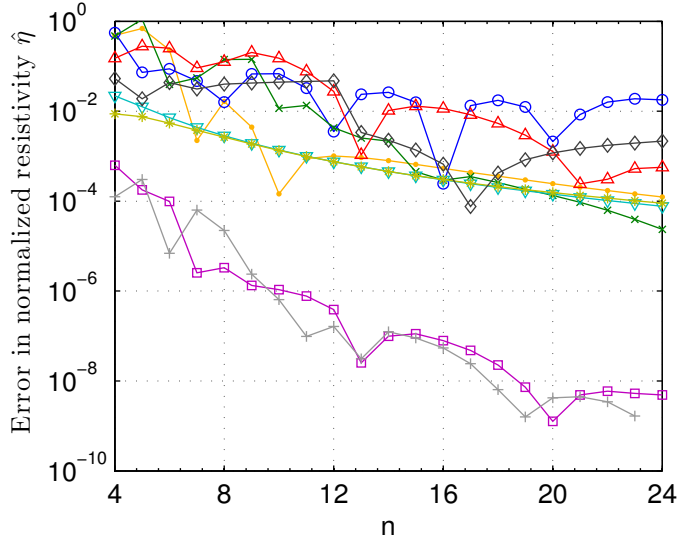}
\caption{(Color online)
Same as figure \ref{fig:resistivity} but computed
using the full linearized Fokker-Planck collision operator.
\label{fig:resistivityFokkerPlanck}}
\end{figure}

\section{Rosenbluth potentials}
\label{sec:potentials}

We now consider issues related to the linearized Fokker-Planck collision operator\cite{RMJ},
which arises in many kinetic theory problems including neoclassical and turbulence calculations.
For many problems of interest, the distribution function $f_a$ of each species $a$ is nearly Maxwellian:
$f_a = f_{a0} + f_{a1}$ with $f_{a1} \ll f_{a0}$ where $f_{a0} = n_a (\pi^{3/2} v_a^3)^{-1} \exp(-x_a^2)$,
$n_a$ is the density,
$v_a=\sqrt{2T_a/m_a}$ is the thermal speed, and $x_a = v/v_a$.
In this case, the collision operator $C_{ab}$ in the kinetic equation of species $a$ due to collisions with
species $b$ may be linearized about the Maxwellians,
and it is given by the sum of the test particle part
$C_{ab}\{f_{a1},f_{b0}\}$ and the field particle part
$C_{ab}\{f_{a0},f_{b1}\}.$
The test particle part is\cite{RMJ, PerBook, BoDarin}
\begin{eqnarray}
\label{eq:Ctest}
C_{ab}\{f_{a1},f_{b0}\}
&=& \Gamma_{ab} n_b \left[ \frac{\erf(x_b)-\Psi(x_b)}{v^3} \LoGyro\{f_{a1}\} \right. \\
&& \left. + \frac{1}{v^2}\frac{\partial}{\partial v}  \left(
2 \frac{m_a}{m_b}\Psi(x_b) x_b^2 f_{a1}
+ \Psi(x_b) v \frac{\partial f_{a1}}{\partial v} \right), \nonumber
\right]
\end{eqnarray}
where $\Gamma_{ab} = 4\pi Z_a^2 Z_b^2 e^4 \ln\Lambda/m_a^2$, $Z_a e$ is the charge of species $a$,
\begin{equation}
\LoGyro=\Lo
+ \frac{1}{2(1-\xi^2)} \frac{\partial^2}{\partial \varphi^2}
\end{equation}
is the Lorentz operator, and the gyrophase $\varphi$ is
the cylindrical angle in velocity space.
Here and for the rest of this paper, $\xi = \vpar/v$ where $\vpar = \vect{v}\cdot\vect{B}/|\vect{B}|$ is
now the component of the velocity along the magnetic field.
The field particle part is
\begin{equation}
C_{ab}\{f_{a0},f_{b1}\} = \Gamma_{ab} f_{a0} \left[
\frac{2v^2}{v_a^4} \frac{\partial^2 G}{\partial v^2} + \frac{2v}{v_a^2} \left( \frac{m_a}{m_b}-1\right) \frac{\partial H}{\partial v}
-\frac{2}{v_a^2} H + 4\pi\frac{m_a}{m_b} f_{b1}
\right]
\label{eq:Cfield}
\end{equation}
where $H$ and $G$ are
the perturbations to the Rosenbluth potentials of species $b$, defined
by
\begin{eqnarray}
\gradv^2 H &=& -4\pi f_{b1} \label{eq:Hdef}\\
\gradv^2 G &=& 2H\label{eq:Gdef}
\end{eqnarray}
with the
velocity-space Laplacian $\gradv^2 = \vv^{-2}\left[ (\partial/\partial
  v) v^2 (\partial/\partial v) + 2\LoGyro \right]$.
(The concise form of the field operator in (\ref{eq:Cfield})
is derived in Eq. (7) of Ref. \onlinecite{BoDarin}.)
For this paper we focus on drift-kinetic calculations,
so we need not retain the gyrokinetic modifications to the test particle\cite{CattoTsang}
and field particle\cite{BoDarin} operators that arise when the collision
operator is gyroaveraged at fixed guiding center position.
We also specialize to the usual case in which only the gyrophase-independent distribution function is considered.
(The gyrophase-dependent part of the distribution function is usually known analytically.)
In this case, we may replace $\LoGyro \to \Lo$.

For the moment, consider a species
affected by collisions only with itself
(such as the ions in a pure plasma),
so we may write the linearized collision operator as
$C_L\{f_{a1}\} = C_{aa}\{f_{a1},f_{a0}\} + C_{aa}\{f_{a0},f_{a1}\}$,
and we drop the species subscripts to simplify notation.
Suppose also we are solving a linear kinetic equation of the form
\begin{equation}
D \{f_1\} + C_L\{f_1\} = R
\label{eq:genericKE}
\end{equation}
where $D$ is some operator describing the kinetic terms and $R$ is some
inhomogeneous term. As we do not know the perturbed Rosenbluth potentials associated
with $f_1$, we may think of (\ref{eq:genericKE}) as having the block matrix structure
\begin{equation}
\left(
\begin{array}{ccc}
M_{11} & M_{12} & M_{13} \\
M_{21} & M_{22} & 0 \\
0 & M_{32} & M_{33} \end{array}
\right) \left(
\begin{array}{c}
f_1 \\
H \\
G \end{array}
\right) = \left(
\begin{array}{c}
R \\
0 \\
0 \end{array}
\right).
\label{eq:blocks}
\end{equation}
The first row in this system corresponds to the original kinetic equation
(\ref{eq:genericKE}).  The second row corresponds to the
Poisson equation (\ref{eq:Hdef}), and the third row similarly corresponds
to (\ref{eq:Gdef}).
The block structure (\ref{eq:blocks}) was pointed out previously in Ref. \onlinecite{Lyons}.
There is no $M_{23}$ or $M_{31}$ element in (\ref{eq:blocks})
since $G$ does not appear in (\ref{eq:Hdef}) and $f_1$ does not appear in (\ref{eq:Gdef}).
We may identify $M_{11}$ with the operator $D$ in (\ref{eq:genericKE}), plus the test-particle collision terms, and also the
last term in (\ref{eq:Cfield})
(which does not involve $H$ or $G$.) Also, $M_{12}$ consists of the $H$ and $\partial H/\partial v$
terms in (\ref{eq:Cfield}), $M_{13}$ is the $G$
term in (\ref{eq:Cfield}), $M_{21} = 4\pi$, $M_{22} = \gradv^2$ with appropriate boundary conditions for $H$,
$M_{32} = -2$, and $M_{33} = \gradv^2$ with appropriate boundary conditions for $G$.

For several reasons, it is beneficial to discretize the $\xi$ dependence of $H$ and $G$
using Legendre polynomials $p_\ell(\xi)$:
$H(z,\xi) = \sum_\ell H_\ell(z) p_\ell(\xi)$ and
$G(z,\xi) = \sum_\ell G_\ell(z) p_\ell(\xi)$
where $z$ represents any other independent variables.
The reasons stem from the fact that the Legendre polynomials
are eigenfunctions of $\gradv^2:$
\begin{equation}
\gradv^2 \left[ W(v) p_\ell(\xi) \right] = p_\ell(\xi) \left[ \frac{1}{v^2} \frac{d}{d v} v^2 \frac{dW}{dv} - \frac{\ell(\ell+1)}{v^2} W(v) \right]
\end{equation}
for any $W(v)$. First, the Legendre representation is extremely efficient, since $\gradv^2 \sim \ell^2$
so $H_\ell \sim 1/\ell^2$, and therefore very few modes need to be retained.
Second, the Legendre discretization allows
efficient treatment of the boundary at large $v$,
which can be understood as follows.
The distribution function will be as small as machine precision for $x>6$, so
it is inefficient to locate any grid points
beyond this value of $x$.
However, $H$ and $G$ vary not as $e^{-x^2}$ but rather as powers of $x$ for large $x$ (as we will show in a moment),
so they are not small on the scale of machine precision even for $x> 6$.
In fact, $|G|$ generally increases to infinity with $x$, so
it is invalid to impose any Dirichlet or Neumann boundary condition on $G$ at large finite $x$.
However, using the Legendre expansion above,
for $x > \xMax=4-6$, the defining equation for $H$ becomes $\gradv^2 H=0$,
and so $H_\Leg =A_\Leg v^{-(\Leg+1)} + B_\Leg v^\Leg$.
The $B_\Leg \ne 0$ solutions are not physical,
and so the Robin boundary condition $\vv \, d H_{\Leg}/d\vv + (\Leg+1) H_{\Leg}=0$ may be applied at $\xMax$ to ensure
the correct behavior $H_{\Leg}\propto v^{-(\Leg+1)}$.
For the $G$ potential, the defining equation
$\gradv^2 G = 2H$ admits
four linearly independent solutions. Two are the homogeneous
solutions to $\gradv^2 G =0$, and the other two are
particular solutions, varying as the homogeneous solutions times $v^2$.
Thus, $G_\Leg =  C_\Leg v^{-(\Leg+1)} + D_\Leg v^\Leg + E_\Leg v^{1-\Leg} + F_\Leg v^{\Leg+2}$.
As can be seen in Eq. (45) of Ref. \onlinecite{RMJ},
the $D_\Leg \ne 0$ and $F_\Leg \ne 0$ solutions are unphysical.
To retain both of the remaining two solutions,
the boundary condition must be second order, and can be found as follows.
Writing $v^2 \,d^2G_\Leg/dv^2 + \beta v \,dG_\Leg/dv + \sigma\,  G_\Leg =0$,
the requirements that $G_\Leg \propto  v^{-(\Leg+1)}$ and $G_\Leg \propto  v^{1-\Leg}$ be solutions
yield two equations for $(\beta,\sigma)$. Thus,
the condition
$\vv^2 \, d^2 G_{\Leg}/d\vv^2 + (2\Leg+1)\vv \, dG_{\Leg}/d\vv + (\Leg^2-1)G_{\Leg}=0$
is the one we impose at $\xMax$.

Legendre polynomials tend to be efficient for representing $f_1$ as well, although for some applications
such as bounce-averaged codes, other representations of the pitch-angle dependence of $f_1$ may be
necessary and/or preferable\cite{Cordey, Lyons, JeffParker}.  For simplicity, we
assume for the rest of this discussion
that the $\xi$ dependence of $f_1$ is discretized in the same Legendre representation as the potentials.
If a different discretization is used for the $\xi$ dependence of $f_1$, $M_{12}$, $M_{13}$, and $M_{21}$
must contain the appropriate matrices for mapping between the representations.

It is much faster to invert the $M_{22}$ and $M_{33}$ blocks of (\ref{eq:blocks})
than to invert $M_{11}$, for two reasons.
First, it typically takes many more Legendre modes to represent $f_1$ than to represent
the potentials. Secondly, the kinetic operator $D$ included in $M_{11}$
generally involves coupling between real space and velocity space coordinates,
whereas $M_{22}$ and $M_{33}$ are independent of real space. In fact, since
$M_{22}$ and $M_{33}$ are diagonal in
the Legendre index, the only coordinate with off-diagonal terms in these blocks is $x$.
Thus, it can be convenient to formally solve the system (\ref{eq:blocks}),
yielding the form
\begin{equation}
\left( M_{11} - \left[ M_{12} - M_{13} M_{33}^{-1} M_{32} \right] M_{22}^{-1} M_{21} \right) f_1 = R.
\label{eq:reduced}
\end{equation}

The fact that $M_{22}$ and $M_{33}$ are very sparse and fast to invert means that extremely
high resolution in $x$ can be used
for $H$ and $G$ for almost no extra cost.  This result is fortuitous, because
the spectral collocation method described in the previous section cannot be used for $H$ or $G$.
This is because, as discussed earlier, the potentials behave as powers of $x$ rather than as $\exp(-x^2)$
for large $x$. Consequently, $H$ and $G$ may be discretized using a very high resolution uniform grid in $x$
with finite-difference derivatives used in $M_{22}$, $M_{33}$, and $M_{13}$.
Or, a high resolution Chebyshev grid in $x$ could be used
for the potentials, with spectral differentiation matrices used in $M_{22}$, $M_{33}$, and $M_{13}$.
(For applications in following sections, we use a uniform grid.)
In either case, the high resolution grid no longer appears once
the matrix multiplications in (\ref{eq:reduced}) are performed,
so the high resolution does not affect the time to solve the system (\ref{eq:reduced}).

A complication introduced by the use of different $x$ grids for $f_1$ and the potentials
is that functions must be mapped from one grid to the other in $M_{12}$, $M_{13}$, and $M_{21}$.
Mapping from the polynomial grid to the uniform grid can be done with spectral accuracy
by evaluating the new polynomials on the uniform grid.
An efficient algorithm for this purpose is the \emph{polint} subroutine from Refs. \onlinecite{DMSuite, DMSuite2, DMSuite3}.
To map from the uniform grid to the polynomial grid, we may use cubic interpolation.
The limited accuracy of this interpolation method is not a problem because
the grid resolution for the potentials can be extremely high at little cost.
Or if the potentials are discretized on a Chebyshev grid, the \emph{polint} subroutine
may again be used to interpolate with spectral accuracy.
To form $M_{12}$, $M_{13}$, and $M_{21}$, both the ``$f_1\to$potentials" and ``potentials$\to f_1$" interpolations
must be expressed as explicit matrices.
As a specific example of these concepts, consider that the matrix $M_{13}$, which corresponds
to the $G$ term in (\ref{eq:Cfield}), may be constructed as
$M_{13} = E r \Delta$ where
$E=\Gamma_{ab}f_{a0} (2v^2/v_a^4)$ is a diagonal operator on the $f_1$ grid,
$r$ is the matrix representing cubic interpolation from the potential grid to $f_1$ grid,
and $\Delta$ is the finite-difference matrix representation of $\partial^2 /\partial x^2$ on the potential grid.

Figure \ref{fig:resistivityFokkerPlanck} shows the plasma resistivity calculation of section \ref{sec:resistivity}
repeated using the full linearized Fokker-Planck operator for electron-electron collisions.
Electron-ion collisions, represented by the $Z$ term in (\ref{eq:Cmod}), are also included.
The number of points in the grid for the Rosenbluth potentials is set to $15 n$ to ensure
it is not a limiting factor in convergence. (Higher resolution in the potential grid causes negligible difference.)
For the Laguerre, associated Laguerre, and Chebyshev methods, the relevant spectral
interpolation scheme is used to interpolate from the polynomial grid to the
uniform grid for the Rosenbluth potentials, just as for the new polynomial scheme.
The efficiency of the new polynomial scheme is again evident.

\section{Application 2: Bootstrap current}

For a more sophisticated application of all the aforementioned numerical techniques, we consider a
calculation of the bootstrap current, a current $\vect{j}$ driven along the magnetic field $\vect{B}$ due
to radial gradients of density and temperature in a toroidal plasma.
In an axisymmetric (tokamak) plasma consisting of electrons and a single ion species, the
bootstrap current has the form\cite{HintonHazeltine, PerBook, Sauter}
\begin{equation}
\left< \vect{j}\cdot \vect{B}\right> =  -cI\pe \left[
  \Lone \frac{1}{\pe}\left(\frac{d\pe}{d\psi} +
  \frac{d\ppi}{d\psi}\right) + \frac{\Ltwo}{\Te}\frac{d\Te}{d\psi} +
  \frac{\Lfour \alpha}{Z \Te} \frac{d\Ti}{d\psi}\right]
  \label{eq:jbs}
\end{equation}
using notation consistent with Ref. \onlinecite{Sauter}.
In (\ref{eq:jbs}),
$c$ is the speed of light,
$I=RB_\zeta$ is the major radius $R$ times the toroidal magnetic field $B_\zeta$, $Z$ is the ion charge,
$\Ti$ and $\Te$ are the ion and electron temperatures, $\ppi$ and $\pe$ are the ion and
electron pressures, and $2\pi\psi$ is the poloidal magnetic flux.
The quantity $\alpha$ is a number obtained by solving the ion kinetic equation, and
the quantities $\Lone$, $\Ltwo$, and $\Lfour$ are moments of the electron
distribution function for various driving terms,
with each distribution obtained by solving the electron kinetic equation.
Specifically,
\begin{equation}
\Lone = -4 \pi^{-1/2} \lla \hat{B} \int _{-1}^1d\xi \int_0^\infty dx\; x^3\xi \feone^{(p)} \rra
\end{equation}
where $\feone^{(p)}$ is the solution of
the normalized electron ``drift-kinetic" equation
\begin{equation}
x \xi \frac{\partial \feone^{(p)}}{\partial\theta}
-\frac{(1-\xi^2)x}{2\hat{B}}\left(\frac{\partial\hat{B}}{\partial\theta}\right)\frac{\partial \feone^{(p)}}{\partial\xi}
- \frac{\nu'}{q R_0 \vect{b}\cdot\nabla\theta} \hat{C} =   \frac{1+\xi^2}{2}x^2 e^{-x^2} \frac{1}{\hat{B}^2} \frac{\partial \hat{B}}{\partial\theta}
\label{eq:fp}
\end{equation}
where $x=x_{\mathrm{e}}$, and the independent variables are $x$, $\xi$, and the poloidal angle $\theta$.
Here,
$\vect{b} = \vect{B}/B$, $B=|\vect{B}|$, $\hat{B} = B/B_0$, $B_0 = I/R_0$,
$R_0$ is a typical major radius, $q$ is the safety factor,
$\nu' = \nue q R_0/\ve$ is a dimensionless
measure of collisionality,
and $\hat{C} = \nue^{-1} [ \Cee + \Cei ]$.
Angle brackets denote a flux surface average:
\begin{equation}
\left< Y \right> = (V')^{-1} \int_0^{2\pi} d\theta\, Y/\vect{B}\cdot\nabla\theta
\end{equation}
for any quantity $Y$ where
\begin{equation}
V' = \int_0^{2\pi} d\theta/\vect{B}\cdot\nabla\theta = \oint
d\ell_\theta / B_\theta,
\label{eq:fsa2}
\end{equation}
$d\ell_\theta$ is the poloidal length element, $V'=dV/d\psi$, and
$2\pi V(\psi)$ is the volume enclosed by the flux surface.

By expanding $\Cei$ in $\sqrt{\me/\mi} \sim 1/60$ for deuterium,
it can be shown that the electron-ion collision operator is dominated\cite{HintonHazeltine, PerBook}
by the $\Lo$ term in the test particle operator (\ref{eq:Ctest}).
Then since $\erf(x)-\Psi(x) \to 1$ for $x\gg 1$, the collision operator
to use in (\ref{eq:fp}) is
$\hat{C}\{\feone^{(p)}\} = \nue^{-1} [ \Cee\{\feone^{(p)}, \feo\} +  \Cee\{ \feo, \feone^{(p)}\} + \Gamma_{\mathrm{ei}} \ni v^{-3} \Lo\{ \feone^{(p)}\} ]$.
As the non-Maxwellian ion distribution function $\fione$ and
perturbed ion Rosenbluth potentials are then no longer present in (\ref{eq:fp}),
it is accurate to within $\sim2\%$
to solve the single-species electron equation (\ref{eq:fp})
without coupling to the ion equation.
If it is desired to retain the full coupling between electrons and ions without
expanding in $\sqrt{\me/\mi}$,
the methods of section \ref{sec:impurities} can be employed.

The other coefficients in the bootstrap current are calculated in a similar manner:
\begin{equation}
\Ltwo = -4\pi^{-1/2}\lla \hat{B} \int _{-1}^1d\xi \int_0^\infty dx\; x^3\xi \feone^{(\Te)} \rra
\end{equation}
where $\feone^{(\Te)}$ is the solution of (\ref{eq:fp})
with an extra factor $x^2-5/2$ multiplying the right-hand side,
and
\begin{equation}
\Lfour = 4\pi^{-1/2}
\lla \hat{B}^2\rra^{-1}
\lla \hat{B} \int _{-1}^1d\xi \int_0^\infty dx\; x^3\xi \feone^{(\Ti)} \rra
\end{equation}
where $\feone^{(\Ti)}$ is the solution of (\ref{eq:fp})
with the right-hand side replaced by
$(3\xi^2-1)(x^2/2)e^{-x^2}\, \partial\hat{B}/\partial\theta$.

We now demonstrate the solution of (\ref{eq:fp}) and the corresponding
equations for $\Ltwo$ and $\Lfour$ using the velocity discretizations described
in the preceding sections.
Some model must be supplied for the $\theta$ dependence of
the dimensionless quantities $\hat{B}$ and $q R_0 \vect{b}\cdot\nabla\theta$.
For this example we use the Miller equilibrium model \cite{Miller} detailed in the appendix.

The $\theta$ coordinate may be discretized in several ways.
In decreasing level of sparsity, some options are finite differencing,
a Fourier modal expansion, or Fourier spectral collocation\cite{Trefethen}.
(The modal approach need not be fully dense in $\theta$ since many coupling terms will be exponentially
small and can therefore be set to zero.)  For results shown here we use the modal approach.
The finite-difference and collocation
approaches lead to simpler code, since multiplications of the unknown fields by
functions of $\theta$ may then be done pointwise, whereas
the modal approach requires matrix multiplication.

If a Gaussian grid is used there is no point at $x=0$, or if the Gauss-Radau grid is used,
there is a grid point at $x=0$.  In the former case there is no boundary
condition imposed at $x=0$,
and in the latter case, the boundary condition $\partial f_{\mathrm{e}1}^{(j)}/\partial x=0$
at $x=0$ is imposed for the $\ell=0$ Legendre mode.
We find results are nearly identical for the two approaches.
Results shown below use the grid with no point at $x=0$.
No boundary condition needs to be applied to $f_{\mathrm{e}1}^{(j)}$ at $x=\infty$,
since the $e^{-x^2}$ behavior is automatically
enforced by the choice of basis functions.

The reduced system (\ref{eq:reduced})
may be solved efficiently using a sparse direct solver, or when the resolution
is sufficiently high, faster solution
is typically possible
using an iterative Krylov-space solver.
For the iterative solvers, preconditioning is essential.  We find an effective preconditioner can be obtained by eliminating
all off-diagonal blocks in the $x$ coordinate.
Calculations shown here are performed using the BICGstab(l)
algorithm \cite{BICGstabl}.

\begin{figure}
\includegraphics[width=\textwidth]{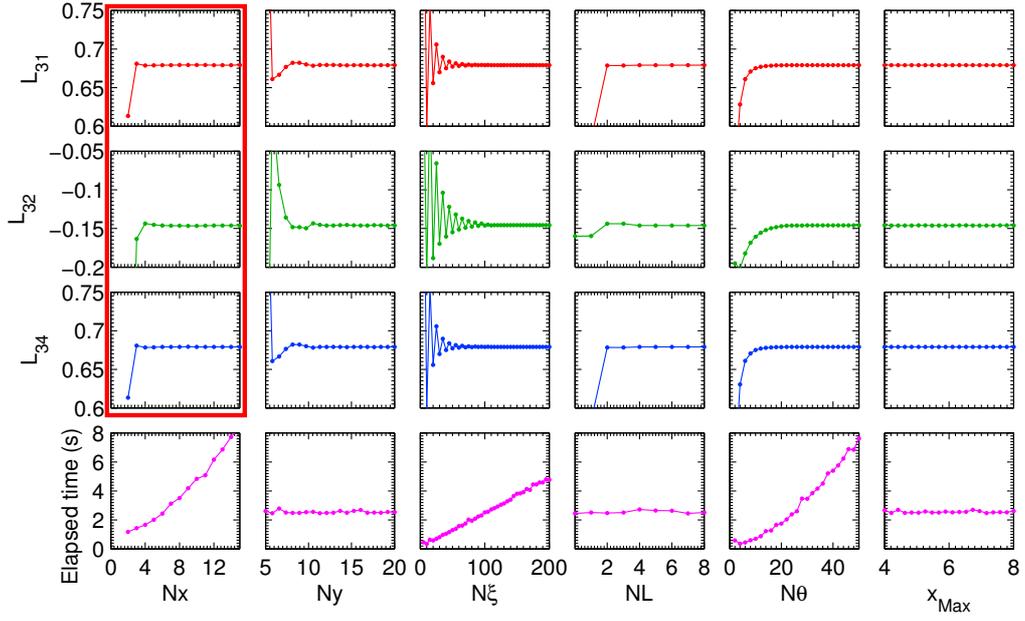}
\caption{(Color online)
Convergence of the bootstrap
current coefficients $\Lone$, $\Ltwo$, and $\Lfour$ in (\ref{eq:jbs})
from the single-species drift-kinetic code
with respect to numerical parameters:
$Nx=$ number of grid points in $x$ for the polynomial grid,
$Ny=$ number of grid points on the uniform $x$ grid for the Rosenbluth potentials,
$N\xi =$ number of Legendre modes used to represent the distribution function,
$NL=$ number of Legendre modes used to represent the potentials,
$N\theta=$ number of Fourier modes in $\theta$,
and $\xMax=$ cutoff of grid for the potentials.
Except for quantities being scanned, the numerical parameters
are $Nx=6$, $Ny=20$, $N\xi=100$, $NL=4$, $N\theta=25$, and $\xMax=5$.
The convergence at a very small $Nx$ enabled by the new discretization scheme
described here is a central
result of this paper.  The collisionality is $\nu_*=0.01$.
\label{fig:singleSpeciesConvergence}}
\end{figure}

Figure \ref{fig:singleSpeciesConvergence} shows the convergence of the code outputs $\Lone$, $\Ltwo$, and $\Lfour$
with respect to the various
numerical parameters,
using the new polynomial grid with no point at $x=0$.  For this run we choose the
collisionality $\nu_* = \nu' A^{3/2}$ to be 0.01, where $A=3.17$ is the aspect ratio.
The excellent convergence with respect to all coordinates is evident.
For the scan of $\xMax$, $Ny$ is varied proportional to $\xMax$ to maintain
constant resolution.
The very rapid convergence with the grid size
in $x$ is enabled by the efficient spectral methods described in the preceding sections.
Convergence to two digits of precision is plenty for comparison with experiment or
for experimental design,
and the code is converged beyond this level with only 4-5 grid points in $x$.
The code execution time is nearly independent of $Ny$, the resolution of the uniform $x$ grid for the potentials,
for the reasons discussed in section \ref{sec:potentials}.
Fortran/Petsc\cite{petsc-web-page, petsc-user-ref} and Matlab versions of the code run in comparable time, since
the rate-limiting step is $LU$-factorization of the preconditioner, which is highly optimized in Matlab.
Execution times reported in Figure \ref{fig:singleSpeciesConvergence} are for the
Matlab version running on a Dell Precision laptop with Intel Core i7-2860 2.50 GHz CPU
and 16 GB memory.

Notice how few Legendre
modes ($\sim 2$) need to be retained in the potentials for good convergence, whereas many more
Legendre modes ($\sim 100$) are required for the distribution function.
The fact that much higher resolution is needed in the $\xi$ and $\theta$ coordinates
than in the $x$ coordinate for $\nu_* \ll 1$ is a result of a boundary layer
known to exist in the solution\cite{HR},
illustrated in Figure 5 of Ref. \onlinecite{usPedestal}.
This sharp feature lies along the curve in the $(\theta,\xi)$ plane
corresponding to the boundary between trapped and circulating particle orbits.
The boundary layer widens as collisionality is increased,
so for collisionalities higher than the small value used here, many fewer modes
in $\theta$ and $\xi$ are required, and execution time is then well below 1 second on a laptop.

Further discussion of code benchmarking and results can be found in Ref. \onlinecite{usPedestal}.

\section{Multiple species}
\label{sec:impurities}

\begin{figure}
\centering
\includegraphics{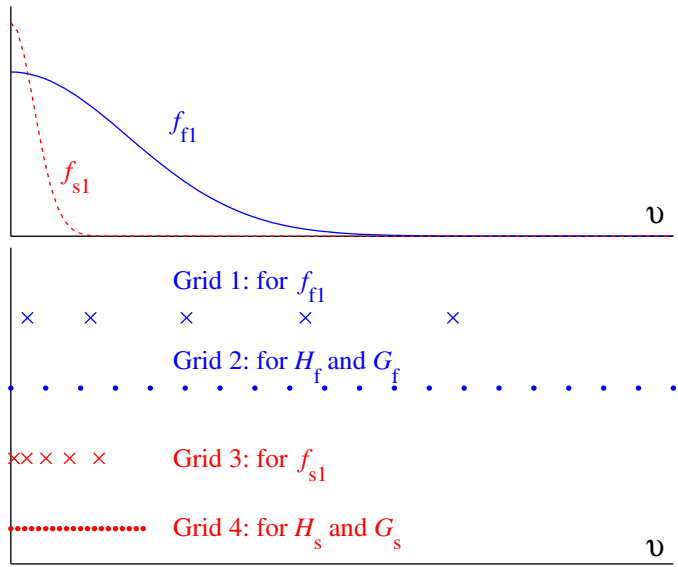}
\caption{(Color online)
For calculations involving $S$ species,
$2S$ grids in the dimensional speed $v$ are required: one for each distribution function,
plus one for the Rosenbluth potentials of each species.
The polynomial grids for the distribution functions
and the uniform grids for the potentials are each
scaled to the thermal speed of the associated species.
Here, grids are shown for a slow species (subscript $s$)
and a fast species (subscript $f$).
\label{fig:multiSpeciesGrids}}
\end{figure}

In computing the bootstrap current, the inter-species coupling was formally negligible,
but for other applications it is essential to solve kinetic equations for multiple species simultaneously,
retaining cross-species interaction.
The thermal speeds of different species may be quite different,
so a grid in $v$ which is appropriate for one species may not be capable of
resolving another species.
To maintain the high accuracy discussed in Section \ref{sec:polynomials},
the distribution function of each species $f_{a1}$
should be represented on the zeros of the $P_n(x_a)$ using
the normalized speed $x_a$ for that particular species,
as shown in figure \ref{fig:multiSpeciesGrids}.
Then the field term (\ref{eq:Cfield}) in the collision operator will require mapping between the $x_a$ grid and $x_b$ grid
and vice-versa.
(The test particle term only involves $f_{b0}$, not $f_{b1}$, so no remapping of unknown functions is required,
and the test particle operator leads to a matrix that is diagonal in species.)
Let us consider how to accurately perform these mappings.

The last term in (\ref{eq:Cfield}), involving $f_{b1}$ and not the potentials, requires mapping
the Maxwellian-like function $f_{b1}$ from the $x_b$ grid to the $x_a$ grid.
This corresponds to mapping between grids 1 and 3 in figure \ref{fig:multiSpeciesGrids}.
This mapping may be accurately done using the \emph{polint} algorithm of Refs.
\onlinecite{DMSuite, DMSuite2, DMSuite3}, as discussed earlier for mapping from the polynomial
grid to the uniform grid.  This algorithm works very well both for evaluating
values of $f_{b1}$ close to $x_b=0$ (which is necessary when $v_b \gg v_a$, mapping a function known on grid 1 to grid 3)
and also for extrapolating to large values of $x_b$ (which is necessary when $v_b\ll v_a$, mapping a function known on grid 3 to grid 1.)
Extrapolation is accurate because functional behavior $\exp(-x_b^2)$ for large $x_b$ is
naturally incorporated into the \emph{polint} algorithm.

Now consider the Rosenbluth potential terms in (\ref{eq:Cfield}).
When $v_b \gg v_a$, the potentials must be evaluated close to
$x_b=0$.
This corresponds to mapping a function known on grid 2 to grid 3 in figure \ref{fig:multiSpeciesGrids}.
Cubic interpolation is effective for this mapping, just as in the single-species
case.  However, when $v_b \ll v_a$, we must extrapolate the potentials to evaluate them beyond
the grid on which they are represented.
This corresponds to mapping a function known on grid 4 to grid 1.
As discussed in section \ref{sec:potentials}, the $H$ potential behaves
as $\propto 1/x_b^{\ell+1}$ for $x_b\to\infty$.
We find in practice it can be very inaccurate to use the simplistic extrapolation $H=0$ for $x_b > \xMax$.
(In particular, the code described in the following section does not converge well
if this extrapolation method is used.)
However, the more accurate extrapolation $H_\ell(x_b) = (\xMax/x_b)^{\ell+1} H_\ell(\xMax)$ for $x_b > \xMax$
works well and leads to good convergence, as
we will demonstrate in the next section.

When $v_b \ll v_a$, we must also extrapolate $d^2 G/dx_b^2$
to form the first left-hand-side term in (\ref{eq:Cfield}).
Notice from Section \ref{sec:potentials} that for large $x_b$, each Legendre mode $\ell$
behaves like
\begin{equation}
\frac{d^2 G_\ell}{dx_b^2} \approx K_\ell x_b^{-\ell-3} + L_\ell x_b^{-\ell-1}
\label{eq:Glimit}
\end{equation}
for constants $K_\ell$ and $L_\ell$.
It is tempting to make the further approximation
$d^2 G_\ell/dx_b^2 \approx L_\ell x_b^{-\ell-1}$,
suggesting the extrapolation
$d^2 G_\ell/dx_b^2  = (\xMax/x_b)^{\ell+1} \left[d^2 G_\ell/dx_b^2\right]_{\xMax}$ for $x_b > \xMax$.
However, since the $K_\ell$ term is only algebraically small compared to the $L_\ell$ term,
this extrapolation requires a much higher $\xMax$ (and therefore many more points in the uniform grid for the potentials)
than a more careful extrapolation which accounts for both the $K_\ell$ and $L_\ell$ terms.
To derive such an extrapolation, we write (\ref{eq:Glimit}) at
the last and penultimate points in the uniform grid for the potentials,
which we denote by $x_{bi}$ and $x_{bj}$ respectively.
Eliminating $K_\ell$ and $L_\ell$ by these two equations, we obtain
the improved extrapolation
\begin{eqnarray}
\frac{d^2 G_\ell}{dx_b^2} &=& \left[\frac{x_{bi}^{\ell+1}}{(x_{bi}^{-2}-x_{bj}^{-2})x_b^{\ell+3}} + \frac{x_{bi}^{\ell+3}}{(x_{bi}^2-x_{bj}^2) x_b^{\ell+1}}\right] \left[\frac{d^2 G_\ell}{dx_b^2}\right]_{x_{bi}} \\
&& - \left[ \frac{x_{bj}^{\ell+1}}{(x_{bi}^{-2}-x_{bj}^{-2})x_b^{\ell+3}} + \frac{x_{bj}^{\ell+3}}{(x_{bi}^2-x_{bj}^2) x_b^{\ell+1}} \right] \left[ \frac{d^2 G_\ell}{dx_b^2}\right]_{x_{bj}}. \nonumber
\end{eqnarray}
We find that this result, which we use for calculations presented below,
leads to convergence for much lower $\xMax$ ($\sim 4-5$).

Due to the factor $f_{a0}$ in (\ref{eq:Cfield}),
every term in (\ref{eq:Cfield}) will be exponentially small for large $x_a$,
even if $v_b \gg v_a$.  Therefore the polynomial $x_a$ grid
is
able to accurately resolve the field particle operator when $v_b \gg v_a$.
However, if $v_b \ll v_a$, features in $f_{b1}$ will be mapped to small $x_a$,
leading to sharp structures that may not be well resolved by the $x_a$ grid.
For this reason, more grid points in $x_a$ may be required when the thermal speeds are very different from each other.
Fortunately, as shown in Figure \ref{fig:newPolynomialZeros},
the new polynomials have zeros clustered near $x=0$, so grid resolution is quite good
in this region, much better than with a Laguerre or associated Laguerre grid. As we will show in the following section,
a demonstration for species with very disparate thermal speeds shows
that the number of grid points in $x$ required for good convergence remains quite small.

\section{Application 3: Impurity flow}

As an example application involving multiple species, we consider the neoclassical flows of
the main ions and a non-trace impurity species in a
tokamak plasma.  Due to the low mass of electrons, electrons do not significantly
affect the motion of the ions, so we do not need to include electrons in the calculation,
though nothing prevents us from retaining electrons if desired.
The perpendicular flow and the variation of the parallel flow on each flux surface
(the so-called \PS flow)
can be written explicitly, so it is only the average parallel flow on a flux surface
that must be determined from kinetic theory.
Following convention, the specific form of the average we compute is $\lla \vect{V}\cdot \vect{B}\rra$.

For this example calculation, we consider a plasma consisting of
deuterium and $\Mo^{+30}$.  We choose the latter since it is often a plentiful
impurity species in metal-wall tokamaks, and its large
mass ($m_{\Mo} = 48 m_{\Deu} = 96$ a.m.u.) tests the performance
of the numerical method when species have
disparate thermal speeds.
We consider an impurity density $n_{\Mo}/n_{\Deu} = 0.0012$, corresponding
to $\Zeff = 2$, where
$\Zeff = (n_{\Deu} + Z_{\Mo}^2 n_{\Mo}) / (n_{\Deu} + Z_{\Mo} n_{\Mo})$.
Since $\Zeff-1$ is not $\ll 1$, the impurities are not in
the trace limit, and so they will have a significant effect on the main-ion flow.

Contributions to the parallel flows arise from radial gradients in the density and temperature
of each species, as well as from the radial electric field.  For this example, we
will compute the contribution from the density gradients, i.e.
we formally set $dT_{\Deu}/d\psi=0$, $dT_{\Mo}/d\psi=0$, and $d\Phi/d\psi=0$.
(Contributions from these other drives could be superposed later due to the linearity of the kinetic equations.)
We assume the density scale lengths are equal:
$n_{\Deu}/(dn_{\Deu}/d\psi) = n_{\Mo}/(dn_{\Mo}/d\psi) $.  We also take the
temperatures of both species to be equal ($T_{\Deu} = T_{\Mo} = T$).

We normalize both flows with the same factor:
\begin{eqnarray}
\lla \vect{V}_{\Deu} \cdot\vect{B} \rra &=& U_{\Deu}\frac{c I T}{e n_{\Deu}}\frac{d n_{\Deu}}{d\psi} \\
\lla \vect{V}_{\Mo} \cdot\vect{B} \rra &=& U_{\Mo} \frac{c I T}{e n_{\Deu}}\frac{d n_{\Deu}}{d\psi}
\end{eqnarray}
and so our goal is to compute the unknown dimensionless quantities $U_{\Deu}$ and $U_{\Mo}$.
To do so, we first solve the appropriate normalized drift-kinetic equations
\begin{eqnarray}
x_{\Deu} \xi \frac{\partial f_{\Deu 1}}{\partial\theta}
-\frac{(1-\xi^2)x_{\Deu}}{2\hat{B}}\left(\frac{\partial\hat{B}}{\partial\theta}\right)\frac{\partial f_{\Deu 1}}{\partial\xi}
- \frac{\nu'}{q R_0 \vect{b}\cdot\nabla\theta} \frac{C_{\Deu}}{\nu_{\Deu}} \hspace{0.5in}\label{eq:DKEForD}\\
 =   \frac{1+\xi^2}{2}x_{\Deu}^2 e^{-x_{\Deu}^2} \frac{1}{\hat{B}^2} \frac{\partial \hat{B}}{\partial\theta} \nonumber
\end{eqnarray}
and
\begin{eqnarray}
x_{\Mo} \xi \frac{\partial f_{\Mo1}}{\partial\theta}
-\frac{(1-\xi^2)x_{\Mo}}{2\hat{B}}\left(\frac{\partial\hat{B}}{\partial\theta}\right)\frac{\partial f_{\Mo1}}{\partial\xi}
- \frac{\nu'}{q R_0 \vect{b}\cdot\nabla\theta} \frac{C_{\Mo}}{\nu_{\Deu}} \hspace{0.5in}\label{eq:DKEForMo} \\
 =   \frac{m_{\Mo}^2 n_{\Mo}}{Z_{\Mo} m_{\Deu}^2 n_{\Deu}}\frac{1+\xi^2}{2}x_{\Mo}^2 e^{-x_{\Mo}^2} \frac{1}{\hat{B}^2} \frac{\partial \hat{B}}{\partial\theta}, \nonumber
\end{eqnarray}
for the distribution functions $f_{\Deu 1}$ and $f_{\Mo1}$,
where
\begin{eqnarray}
C_{\Deu} &=& C_{\Deu \Deu}\{f_{\Deu 1},f_{\Deu 0}\} + C_{\Deu \Deu}\{f_{\Deu 0},f_{\Deu 1}\}  \\
&& + C_{\Deu \Mo}\{f_{\Deu 1},f_{\Mo0}\} + C_{\Deu \Mo}\{f_{\Deu 0},f_{\Mo1}\}, \nonumber
\end{eqnarray}
and
\begin{eqnarray}
C_{\Mo} &=& C_{\Mo\Mo}\{f_{\Mo1},f_{\Mo0}\} + C_{\Mo\Mo}\{f_{\Mo0},f_{\Mo1}\} \\
&&+ C_{\Mo \Deu}\{f_{\Mo1},f_{\Deu 0}\} + C_{\Mo\Deu}\{f_{\Mo0},f_{\Deu 1}\}. \nonumber
\end{eqnarray}
This time, $\nu' = \nu_{\Deu} q R_0/v_{\Deu}$ in both equations (\ref{eq:DKEForD})-(\ref{eq:DKEForMo}), where
$\nu_{\Deu} = 4\sqrt{2\pi} n_{\Deu} e^4\ln\Lambda/(3\sqrt{m_{\Deu}}T^{3/2})$.
Finally, we compute the moments
\begin{eqnarray}
U_{\Deu} &=& \frac{4}{\pi^{1/2}} \lla \hat{B} \int_0^\infty dx_{\Deu}\, x_{\Deu}^3 \int_{-1}^1 d\xi\, \xi f_{\Deu 1} \rra \\
U_{\Mo} &=& \frac{4}{\pi^{1/2}} \left( \frac{m_{\Deu}}{m_{\Mo}}\right)^2 \frac{n_{\Deu}}{n_{\Mo}}\lla \hat{B} \int_0^\infty dx_{\Mo}\, x_{\Mo}^3 \int_{-1}^1 d\xi\, \xi f_{\Mo1} \rra.
\end{eqnarray}

We again use the Miller geometry detailed in the appendix.
A preconditioning matrix is employed, obtained by dropping not only the blocks that are off-diagonal in $x$,
as in the single-species code, but also dropping all blocks that are off-diagonal in species.

\begin{figure}
\includegraphics[width=\textwidth]{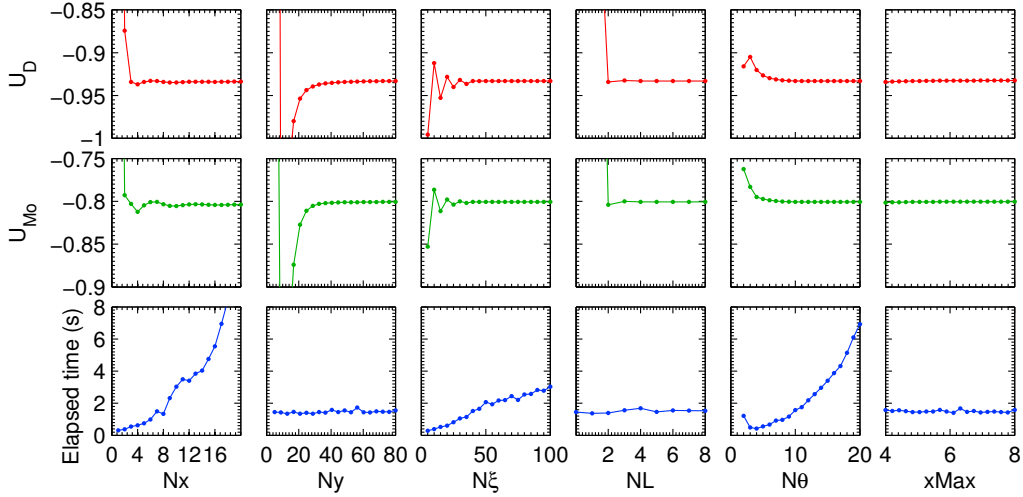}
\caption{(Color online)
Convergence of $U_{\Deu}$ and $U_{\Mo}$ -- the normalized parallel flows of $\Deu^{+1}$ and $\Mo^{+30}$ --
in the multiple-species drift-kinetic code
with respect to numerical parameters defined as in Figure \ref{fig:singleSpeciesConvergence}.
Except for quantities being scanned, the numerical parameters
are $Nx=8$, $Ny=80$, $N\xi=40$, $NL=4$, $N\theta=10$, and $\xMax=5$.
Good convergence at very small values of $Nx$ is again apparent.
The collisionality is $\nu_*=0.01$.
\label{fig:impurityConvergence}}
\end{figure}

Figure \ref{fig:impurityConvergence} shows the convergence of the two-species calculation.
Calculations are performed for $\nu_{*\Deu} = \nu' A^{3/2}= 0.01$.
For simplicity, in each run we use the same resolution for both species.
As for the single-species case,
convergence is achieved at a very small number of grid points in $x$,
with 1\% accuracy achieved by $Nx=5$.
Higher $Ny$ is required, but this causes no decrease in code performance,
as shown in the figure, for the reasons discussed in Section \ref{sec:potentials}.

\section{Discussion}

A common feature of kinetic simulations is the need to accurately evaluate both integrals
and derivatives of Maxwellian-like functions using a single grid or modal discretizaton.
In this work we have presented a set of methods which allow very accurate
integration and differentiation for such functions even with very coarse
grids.  The central concept is to employ polynomial approximation
using the weight $\exp(-x^2)$ and interval $[0,\infty)$, which
is equivalent to an expansion in the new orthogonal polynomials
(\ref{eq:monicPolynomials})-(\ref{eq:normalizedPolynomials}).
The Stieltjes procedure (detailed in Refs. \onlinecite{Gautschi1, Gautschi2, Gautschi3, ORTHPOL}
or Section 4.6.2-3
of Ref. \onlinecite{numericalRecipes}, and summarized in Section \ref{sec:polynomials} of the present paper) is used to generate
the grid and integration weights.
Then the \emph{poldif} and \emph{polint}
routines of Refs. \onlinecite{DMSuite, DMSuite2, DMSuite3} are used
to generate differentiation and interpolation matrices.
These algorithms 
are applicable to any weight function and any integration interval,
and were not developed with velocity-space applications in mind.
The recognition that these algorithms may be applied to kinetic calculations,
using the interval and weight appropriate for the $v$ or $v_\bot$ coordinate of Maxwellian-like
functions, is novel.
Figure \ref{fig:grids1} is a central result of our work,
showing the new method performs far better than other
discretization schemes at both integrating and differentiating
functions relevant to kinetic theory.
The abscissae, integration weights, and differentiation matrices
are not significantly more difficult to compute than
those for classical orthogonal polynomials, and
source code is available for generating the grid, integration weights,
differentiation matrices, and interpolation matrices.
Figure (\ref{fig:resistivity}) shows the success of the new method
in a model physics application -- a calculation of resistivity in a uniform plasma.

We have also shown how the discretization scheme may be applied
when the Fokker-Planck collision operator is included
in kinetic calculations.  The Rosenbluth potentials in the
collision operator cannot be discretized in the same manner
as the distribution function because they do not behave as $\propto \exp(-x^2)$
for large $x$.  Instead, they may be discretized on a uniform or Chebyshev grid with very
high resolution.  By the construction
(\ref{eq:reduced}), this high resolution grid is eliminated through matrix multiplication
when the final matrix for the kinetic equation is constructed,
and so the grid resolution for the potentials does not affect the size
of the final matrix.  Consequently, overall code performance and memory requirements are
essentially independent of the resolution
of the grid for the potentials, and the final matrix size is
determined instead by the much coarser resolution
of the grid for the distribution function.

The methods have been demonstrated in several neoclassical calculations
for a tokamak plasma, which involve the solution of one or more
drift-kinetic equations including collisions.
Figure \ref{fig:singleSpeciesConvergence}
illustrates such a calculation for the bootstrap current,
showing that as few as four grid points in $x$ may be sufficient for convergence
of two significant digits (which is more than enough accuracy for comparison to experiment.)
When species with disparate masses are included simultaneously,
as in the impurity flow computation of figure \ref{fig:impurityConvergence},
careful extrapolation of the Rosenbluth potentials must be performed,
but as long as this is done, rapid convergence is achieved for a similarly
small number of grid points in $x$.


\section*{Acknowledgements}
The authors are grateful to Michael Barnes for assistance regarding the GS2 grid.
  This work was
supported by the
Fusion Energy Postdoctoral Research Program
administered by the Oak Ridge Institute for Science and Education.

\appendix

\section{Miller geometry}
Miller equilibria were defined in Ref. \onlinecite{Miller}.
Additional relations for Miller and other local MHD equilibria,
used in gyrokinetic codes GS2 and GYRO,
are systematically reviewed in Ref. \onlinecite{CandyGeometry}.
For clarity, 
here we present
the relations for Miller equilibria that are
needed for the neoclassical calculations described in the main text. Specifically,
we require
$\vect{B}\cdot\nabla\theta$ and the field magnitude $B$, both as functions
of a poloidal angle $\theta$.

The Miller formulation begins with the assumption that the flux surface of interest has the shape
\begin{eqnarray}
R(\theta) &=& R_0 +   r \cos(\theta+x\sin\theta), \label{eq:R}\\
Z(\theta)&=& \kappa r \sin\theta,
\end{eqnarray}
where $R_0$ is the major radius at the geometric center of the flux surface, $r$ is the minor radius,
$\kappa$ is the elongation,
$x = \arcsin\delta$, and $\delta$ is
the triangularity.  The aspect ratio $A$ may be defined by $A=R_0/r$. All the quantities $R_0$, $r$, $\kappa$, $\delta$, and $A$ are considered
to be functions of the flux surface label $\psi$.
This $\theta$ angle is generally not the angle $\arctan(Z/(R-R_0))$,
nor is it the Boozer or Hamada angle or other straight-field-line coordinate.

Next, consider the standard relation for any $(q^1, q^2, q^3)$ curvilinear coordinates
\begin{equation}
\nabla q^1 = \frac{1}{J}\frac{\partial\vect{x}}{\partial q^2}\times \frac{\partial\vect{x}}{\partial q^3}
\label{eq:generalGradient}
\end{equation}
where
\begin{equation}
J = \frac{\partial\vect{x}}{\partial q^1}
\cdot \frac{\partial\vect{x}}{\partial q^2}\times \frac{\partial\vect{x}}{\partial q^3}
 = \left( \nabla q^1 \cdot \nabla q^2 \times \nabla q^3\right)^{-1}
\label{eq:generalJacobian}
\end{equation}
is the Jacobian, and here we choose $(q^1, q^2, q^3) = (\psi,\zeta,\theta)$ with $\zeta$
the toroidal angle.
Recalling that any axisymmetric field may be written $\vect{B}=\nabla\zeta\times\nabla\psi+I\nabla\zeta$,
the quantity $\vect{B}\cdot\nabla\theta$ we are seeking is equal to $1/J$.
To evaluate the derivatives in (\ref{eq:generalGradient})-(\ref{eq:generalJacobian}),
we write the position vector as
\begin{equation}
\vect{x} = \vect{e}_X R \cos\zeta + \vect{e}_Y R \sin\zeta + \vect{e}_Z Z,
\label{eq:positionVector}
\end{equation}
where $(\vect{e}_X, \vect{e}_Y, \vect{e}_Z )$ are the Cartesian unit vectors.
Differentiating (\ref{eq:positionVector}), we obtain, after some algebra,
\begin{equation}
J = \frac{1}{\vect{B}\cdot\nabla\theta}= \frac{Y}{\partial\psi/\partial r}
\label{eq:Jacobian}
\end{equation}
where
\begin{eqnarray}
Y&=&  \kappa r R \left\{ \sin(\theta+x\sin\theta)\left[ 1+x\cos\theta\right](1+s_\kappa)\sin\theta \frac{}{}\right. \\
&& \left. +\cos\theta\left[ \frac{\partial R_0}{\partial r}+\cos(\theta+x\sin\theta)-s_\delta\sin(\theta+x\sin\theta)\sin\theta\right] \right\},  \nonumber
\end{eqnarray}
$s_\kappa = (r/\kappa)\partial \kappa/\partial r$ and $s_\delta = (1-\delta^2)^{-1/2} r \,\partial\delta/\partial r$.
Then from the magnitude of (\ref{eq:generalGradient}),
\begin{equation}
|\nabla\psi| = \frac{r R}{J}\sqrt{ \left\{ \sin(\theta+x\sin\theta) \left[1+x\cos\theta\right]\right\}^2 + (\kappa\cos\theta)^2}.
\label{eq:gradPsi}
\end{equation}
Defining $B_0 = I/R_0$, the field magnitude may be written
\begin{equation}
B = \sqrt{B_\theta^2 + (B_0 R_0/R)^2}
\label{eq:BMagnitude}
\end{equation}
where $B_\theta = |\nabla\psi|/R = $(\ref{eq:gradPsi})$/R$ is the poloidal field.
The $B_\theta$ calculated in this manner from (\ref{eq:Jacobian})-(\ref{eq:gradPsi}) is precisely Eq. (37) of Ref. \onlinecite{Miller}.

Equations (\ref{eq:Jacobian})-(\ref{eq:BMagnitude}) with (\ref{eq:R})
are the desired expressions for the two quantities we need, $B$ and $\vect{B}\cdot\nabla\theta$, albeit in terms of the
quantity $\partial\psi/\partial r$ in (\ref{eq:Jacobian}).  It is usually preferable to specify the safety factor $q$ in place of $\partial\psi/\partial r$.  To do so, we
observe
\begin{equation}
q = \frac{1}{2\pi}\int_0^{2\pi}d\theta\frac{d\zeta}{d\theta} =  \frac{1}{2\pi}\int_0^{2\pi}d\theta\frac{\vect{B}\cdot\nabla\zeta}{\vect{B}\cdot\nabla\theta}.
\end{equation}
Then using (\ref{eq:Jacobian}),
\begin{equation}
\frac{\partial\psi}{\partial r} = \frac{B_0 R_0}{2\pi q}\int_0^{2\pi}d\theta\frac{Y}{R^2}
\end{equation}
where the integral may be evaluated numerically.

In summary,
the magnetic geometry is fully specified by the following seven dimensionless parameters:
$A$ (aspect ratio),
$\kappa$ (elongation),
$\delta$ (triangularity),
$q$ (safety factor),
$s_\kappa$ (elongation shear),
$s_\delta$ (triangularity shear),
and $\partial R_0 /\partial r$ (related to the Shafranov shift).
Since only the normalized quantities $\hat{B}=B/B_0$ and $q R_0 \vect{b}\cdot\nabla\theta$
appear in the code, there is no need to specify
the two dimensional parameters
$r$ (minor radius) and $B_0$ (magnetic field at the geometric center of the flux surface.)
The additional dimensionless parameters $s$ (global magnetic shear) and $\alpha$ (normalized
pressure gradient) are discussed in Ref. \onlinecite{Miller} but are
not needed for the calculations here.

The calculations for figures \ref{fig:singleSpeciesConvergence}
and \ref{fig:impurityConvergence} were performed using the following
parameters, identical to Ref. \onlinecite{Miller}:
$A=3.17, \kappa=1.66, \delta = 0.416, s_\kappa = 0.70, s_\delta = 1.37, \partial R_0/\partial r = -0.354$,
and $q=3.03$.





\bibliographystyle{model1a-num-names}
\bibliography{speedGrids}







\end{document}